\documentclass[11pt]{iopart}
\pdfoutput=1
\expandafter\let\csname equation*\endcsname\relax
\expandafter\let\csname endequation*\endcsname\relax
\usepackage[numbers,sort&compress]{natbib} 
\usepackage[pdftex]{graphicx,color}
\usepackage{amsmath,amsfonts,amssymb,wasysym,graphicx,capt-of,ifthen,calc}
\usepackage{enumerate,float,url,color}
\usepackage{amsmath}
\usepackage{graphics}  
\usepackage{psfrag} 
\usepackage[latin1]{inputenc}
\usepackage{color}
\usepackage{rotating}
\usepackage{url}
\usepackage{hyperref}
\usepackage{mathtools}
\usepackage[caption=false]{subfig}
\usepackage{bbm}
\begin{document}
\title{Survival probability of random walks leaping over traps}
\author{Gaia Pozzoli}
\address{Dipartimento di Scienza e Alta Tecnologia and Center for Nonlinear and Complex Systems, Universit\`a degli Studi dell'Insubria, Via Valleggio 11, 22100 Como Italy}
\address{I.N.F.N. Sezione di Milano, Via Celoria 16, 20133 Milano, Italy}
\author{Benjamin De Bruyne}
\address{LPTMS, CNRS, Univ.\ Paris-Sud, Universit\'e Paris-Saclay, 91405 Orsay, France}

\begin{abstract}
  We consider one-dimensional discrete-time random walks (RWs) in the presence of finite size traps of length $\ell$ over which the RWs can jump. We study the survival probability of such RWs when the traps are periodically distributed and separated by a distance $L$. We obtain exact results for the mean first-passage time and the survival probability in the special case of a double-sided exponential jump distribution. While such RWs typically survive longer than if they could not leap over traps, their survival probability still decreases exponentially with the number of steps. The decay rate of the survival probability depends in a non-trivial way on the trap length $\ell$ and exhibits an interesting regime when $\ell\rightarrow 0$ as it tends to the ratio $\ell/L$, which is reminiscent of strongly chaotic deterministic systems. We generalize our model to continuous-time RWs, where we introduce a power-law distributed waiting time before each jump. In this case, we find that the survival probability decays algebraically with an exponent that is independent of the trap length. Finally, we derive the diffusive limit of our model and show that, depending on the chosen scaling, we obtain either diffusion with uniform absorption, or diffusion with periodically distributed point absorbers.
\end{abstract}

\section{Introduction}
The study of random walks (RWs) has a long history both for practical applications and from a purely mathematical point of view. One classical problem in the theory of RWs is the trapping of a particle by an environment composed of multiple traps that absorb the particle once they encounter it \cite{Lifshitz63,Lifshitz65,Lifshitz88,Balagurov74,Rosenstock70,Donsker75,Donsker79}. Trapping problems have a long-standing interest in the physics community and beyond as they govern the behavior of a variety of applications ranging from target searching strategies \cite{Oshanin02} to chemical kinetics and diffusion-limited reactions \cite{Smol16,Chand43,Rice85,Montroll(1965),Benson60,Krapivsky10}. Such problems have been studied with various dynamics for the particle and in a wide variety of static, dynamic and random environments \cite{Bramson88,Bray02a,Bray02b,Bray02c,Majumdar03,Bray03,Yuste08,Krapivsky2014,Ledoussal2009,Texier2009,Grabsch2014}. 
From a theoretical point of view, they have shown to exhibit quite a rich behavior with non-trivial features such as a slower-than-exponential decay of the survival probability in the case of randomly distributed traps \cite{Lifshitz63,Lifshitz65,Lifshitz88,Balagurov74,Donsker75,Donsker79}. 

In trapping problems, traps are generally represented by point absorbers, either on a lattice or in continuous space, and their spatial extent  is usually neglected. This assumption eases the analytical treatment and sometimes permits for an exact solution of the survival probability to be found. However, such assumption might not hold for trapping processes where the spatial dimensions of the traps are relevant. One example where the spatial extent of the traps is particularly important is the search of ``non-revisitable" targets, which is of interest in several practical circumstances such as animal foraging \cite{Giuggioli05,Randon09,MajumdarCom10,Murphy92,Boyle09} and time-sensitive rescue missions \cite{Shlesinger(2006)}. The spatial extent of the traps could also be important in other phenomena such as electron-hole recombination on a surface \cite{Klafter(1985)}, risk control and extreme value statistics \cite{Masoliver(2005), Montero(2007)} in mathematical finance \cite{Scalas(2000), Mainardi(2000), Masoliver(2003), Scalas(2006)}, or  diffusion-controlled reactions, where the survival probability turns out to be directly related to the time evolution of the concentration of the chemical species  \cite{Havlin(1987),Blumen(1984)}. In a different but related problem, the mean exit time of a diffusive particle from a bounded domain through a finite size opening is known to exhibit a non-trivial behavior, in particular in the narrow escape limit \cite{Rupprecht15,Metzler14,rednerGuide,Schuss1,Schuss2,Schuss3,Schuss4}.

\begin{figure}[t]
  \begin{center}
    \includegraphics[width=0.6\textwidth]{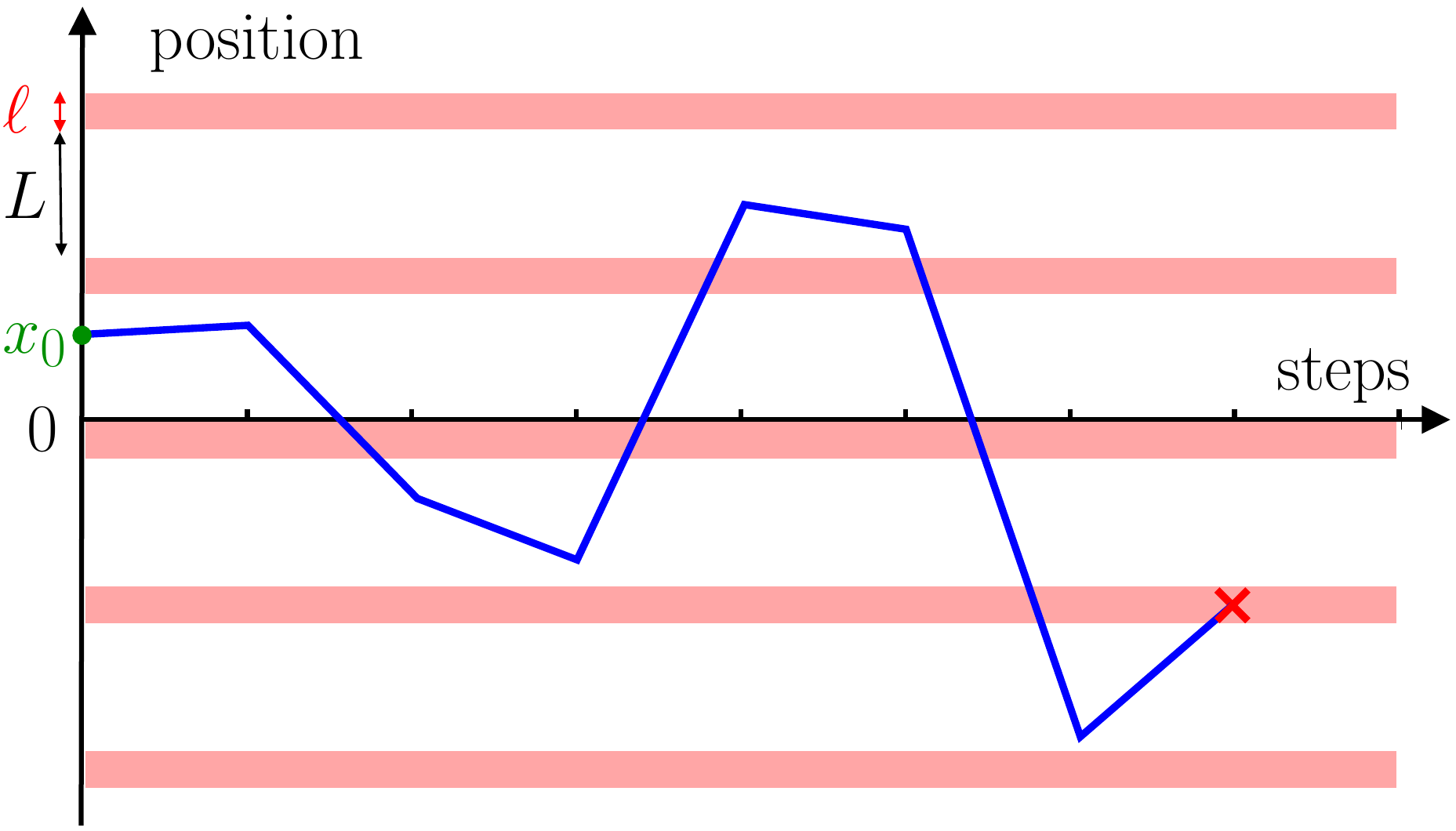}
    \caption{Schematic representation of a trajectory (blue line) of a random walk leaping over periodically distributed traps of length $\ell$ (light red stripes) and separated by a distance $L$. The blue trajectory started at $x_0$ and survived during $7$ steps, before landing \emph{exactly} into one of the traps (red cross). }
    \label{fig:model}
  \end{center}
\end{figure}

A natural question then arises: ``How is the survival probability of a particle affected in the presence of traps with a finite size?''. The main goal of this paper is to answer this question. We study a simple model where the environment is a one-dimensional line with periodically distributed traps of finite length $\ell$ and separated by a distance $L$ (see figure \ref{fig:model}). In this environment, the one-dimensional random walk $x_n$ evolves according to the Markov rule
\begin{align}
  x_n = x_{n-1} + \eta_n\,,\label{eq:mr}
\end{align}
starting from $x_0$, where $\eta_n$'s are i.i.d.~random variables drawn from a distribution $f(\eta)$. The distinctive feature of this model is that the random walk explores the real line while \emph{leaping over} traps, until it eventually jumps \emph{exactly} into one of the traps  (see figure \ref{fig:model}), which is in contrast with the usual absorbing and reflecting barriers.

In this paper, we obtain explicit results on the survival probability of the random walk leaping over finite size traps. For the case of a double-sided exponential jump distribution, we obtain the mean first-passage time [see equation (\ref{eq:mpftsolx})] and show that the survival probability decreases exponentially with the number of steps with a decay rate that depends on the trap size in a non-trivial way [see equations (\ref{eq:Sdecay}), (\ref{eq:detzero}) and the discussion below]. In addition, we generalise our model to the case of continuous-time random walks (CTRWs) with power-law distributed waiting times at each step, for which we obtain an algebraic decay of the survival probability for long times [see equation (\ref{eq:SCTRWs5})]. 
Finally, we derive the diffusive limit of our model and show that, depending on the chosen scaling, we obtain either diffusion with uniform absorption, or diffusion with periodically distributed point absorbers [see equations (\ref{eq:difftrap}) and (\ref{eq:diffab})].

The paper is structured as follows. In Section \ref{sec:dtrw}, we introduce the discrete-time model, which is the starting point for all the subsequent sections. We write a recurrence relation for the survival probability  of general validity and then focus on the solution for an exactly solvable example, for which we obtain an exact expression for the mean first-passage time and the asymptotic decay of the survival probability. In Section \ref{sec:ctrw}, we generalise our model to CTRWs in the presence of fat-tailed waiting times between steps. In Section \ref{sec:dl}, we derive the continuum limit of the discrete-time model. Finally, we provide a summary and perspectives for further research. Some detailed calculations are presented in the Appendices.

\section{Discrete-time random walk model}\label{sec:dtrw}

In this section, we derive a backward equation for the survival probability for the random walk (\ref{eq:mr}), with an arbitrary jump distribution $f(\eta)$. We solve this equation for a particular jump distribution, namely the double-sided exponential distribution, and obtain the mean first-passage time as well as the asymptotic behavior of the survival probability for a large number of steps. 
\subsection{Survival probability}
\label{sec:surv}
\begin{figure}[t]
  \begin{center}
    \includegraphics[width=0.4\textwidth]{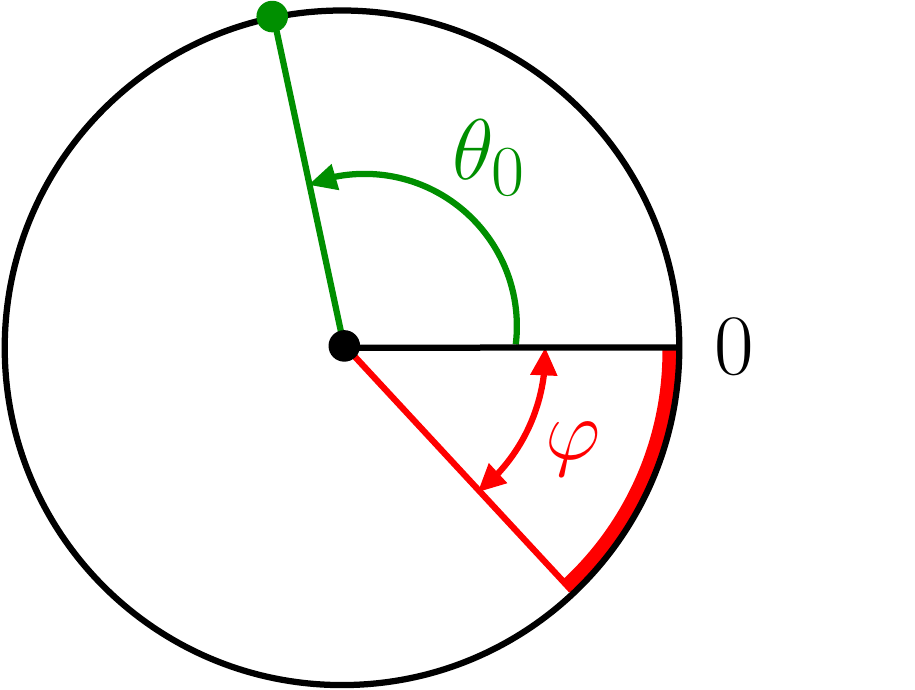}
    \caption{The random walk leaping over periodically distributed traps in figure \ref{fig:model} is equivalent to a random walk on a circle of perimeter $L+\ell$ leaping over a single trap of arc length $\ell$ located on the arc $]2\pi-\varphi,2\pi[$ with $\varphi= \frac{2\pi\,\ell}{L+\ell}$ (red region). The initial position $x_0$ of the random walk is mapped to the initial angular position $\theta_0=\text{mod}\left(\frac{2\pi x_0}{L+\ell},2\pi\right)$.}
    \label{fig:circle}
  \end{center}
\end{figure}

Due to the periodicity of the environment in figure \ref{fig:model}, it is equivalent to study a random walk on a circle of perimeter $L+\ell$ equipped with a single trap of arc length $\ell$ located on the arc $]2\pi-\varphi,2\pi[$ with the angular trap size (see figure \ref{fig:circle})
\begin{align}
  \varphi \coloneqq \frac{2\pi \ell}{L+\ell}\,. \label{eq:varphi}
\end{align} 
Denoting the angular position of the random walk $\theta_n=2\pi\,\text{mod}(x_n/(L+\ell),1)$, the Markov rule (\ref{eq:mr}) becomes
\begin{align}
  \theta_n = \text{mod}\left(\theta_{n-1} + \frac{2\pi\,\eta_n}{L+\ell}\,,\,2\pi\right)\,, \label{eq:eomc}
\end{align}
where $\eta_n$'s are i.i.d.~random variables drawn from a jump distribution $f(\eta)$ and where the initial angular position is given by
\begin{align}
  \theta_0 =\text{mod}\left(\frac{2\pi x_0}{L+\ell}\,,\,2\pi\right)\,. \label{eq:theta0}
\end{align} 
 We would like to compute the survival probability $S_{\varphi}(n\,|\,\theta_0)$ of the random walk  after $n$ steps in the presence of a trap of angular size $\varphi$ given that it started at the angular position $\theta_0$, which is defined by
\begin{align}
  S_{\varphi}(n\,|\,\theta_0) = \text{Prob.}\left(\theta_1 \leq 2\pi-\varphi\,,\, \ldots\,,\,\theta_n \leq 2\pi- \varphi \,|\,\theta_0\right)\,,\label{eq:Sc}
\end{align} 
where $0\leq \theta_0\leq 2\pi-\varphi$ is the initial angular position.
The survival probability satisfies the recursive backward equation
\begin{align}
    S_{\varphi}(n\,|\,\theta) &= \frac{L+\ell}{2\pi}\,\sum_{m=-\infty}^{\infty}\int_{2\pi m}^{2\pi(m+1)-\varphi} d\theta' \, f\left(\frac{(L+\ell)\,(\theta'-\theta)}{2\pi}\right)\, S_{\varphi}(n-1\,|\,\theta')\,, \label{eq:Sreca}
   \end{align}
   where the sum appears due to the modulo operator in the Markov rule (\ref{eq:eomc}) and the prefactor comes from the change of variable in the jump distribution from the step $\eta$  on the real line to the angular step $2\pi \eta/(L+\ell)$ on the circle. By shifting the integration variable by $2\pi m$ and switching the order of the sum and the integral, the recursive equation becomes
   \begin{align}
    S_{\varphi}(n\,|\,\theta)  &= \int_0^{2\pi-\varphi} d\theta' F(\theta'-\theta)\, S_{\varphi}(n-1\,|\,\theta')\,,\label{eq:Srec}
\end{align}
where we used the periodicity $S_{\varphi}(n\,|\,\theta)=S_{\varphi}(n\,|\,\theta+2\pi)$ and defined the periodised angular jump distribution $F(\theta)$ given by
\begin{align}
  F(\theta) = \frac{L+\ell}{2\pi}\sum_{m=-\infty}^{\infty} f\left(\frac{(L+\ell)(\theta + 2\pi m)}{2\pi}\right)\,.\label{eq:Fper}
\end{align}
 Note that, by definition, we have the periodic property $F(\theta+2\pi)=F(\theta)$ and the normalisation property $\int_0^{2\pi}d\theta\, F(\theta-\phi)=1$, for all $\phi$. The integral equation for the survival probability (\ref{eq:Srec}) is valid for an arbitrary jump distribution. However, the integral, which extends only over the interval $[0,2\pi-\varphi]$, is known to be of Fredholm type and is notoriously difficult to solve for an arbitrary jump distribution \cite{Feller,Morse}. Fortunately, there is one exactly solvable case which we present in the next section. 
\subsection{An exactly solvable case}
Let us consider the case where the jump distribution $f(\eta)$ is a double-sided exponential distribution
\begin{align}
  f(\eta) = \frac{1}{\sqrt{2}\sigma}\,e^{-\frac{\sqrt{2}|\eta|}{\sigma}}\,,\label{eq:dse}
\end{align}
where $\sigma^2$ is the variance of the jump distribution. 
Inserting the jump distribution (\ref{eq:dse}) into the periodised version (\ref{eq:Fper}), we find that the periodised jump distribution is
\begin{align}
  F(\theta) = \frac{1}{\sqrt{2}\sigma_a}\,e^{-\frac{\sqrt{2}|\theta|}{\sigma_a}} + \frac{\sqrt{2}}{\sigma_a(e^{2\pi}-1)}\cosh\left(\frac{\sqrt{2}\,\theta}{\sigma_a}\right)\,,\label{eq:Fperdes}
\end{align}
where we introduced the angular standard deviation $\sigma_a$ given by
\begin{align}
  \sigma_a \coloneqq \frac{2\pi\sigma}{L+\ell}\,,\label{eq:sigmaa}
\end{align}
where the subscript $a$ stands for ``angular'' standard deviation. 
One can check that $F(\theta+2\pi)=F(\theta)$ and $\int_0^{2\pi}d\theta\, F(\theta-\phi)=1$, for all $\phi$, which might not be obvious at first sight. For this particular jump distribution, the integral equation (\ref{eq:Srec}) can be solved by using the well-known trick of the double-sided jump distribution (see for instance \cite{Majumdar06}) which consists in taking twice a derivative with respect to $\theta$ of the integral equation (\ref{eq:Srec}) and using the property that
\begin{align}
 \frac{\sigma_a^2}{2}\, F''(\theta) = F(\theta) - \delta(\theta)\,,\label{eq:dsep}
\end{align}
to obtain the differential equation
\begin{align}
    \frac{\sigma_a^2}{2}\,\partial_{\theta\theta} S_{\varphi}(n\,|\,\theta) = S_{\varphi}(n\,|\,\theta) - \Theta(2\pi-\varphi-\theta)\,S_{\varphi}(n-1\,|\,\theta)\,,\quad 0\leq \theta < 2\pi\,,\label{eq:dsediff}
\end{align}
where $\Theta(x)$ is the Heaviside step function with $\Theta(x)=1$ if $x\geq 0$ and  $\Theta(x)=0$ otherwise. Note that the domain where $\theta \in ]2\pi-\varphi,2\pi[$ is not physically relevant as it corresponds to the interior of the trap. In principle, one can solve the differential equation (\ref{eq:dsediff}) on the physically relevant interval $\theta \in [0, 2\pi-\varphi]$ and use the integral equation (\ref{eq:Srec}) to determine the integration constants. Alternatively, one can solve the differential equation on the full circle and apply periodic boundary conditions, and subsequently restrict the solution to the physically relevant interval $\theta \in [0, 2\pi-\varphi]$. We follow the latter approach.
To solve the equation (\ref{eq:dsediff}), it is convenient to consider the generating function of the survival probability
\begin{align}
q_{\varphi}(z,\theta)= \sum_{n=1}^\infty S_{\varphi}(n\,|\,\theta)\,z^n\,,\label{eq:dsegen}
\end{align}
which, upon using equation (\ref{eq:dsediff}), satisfies 
\begin{align}
  \frac{\sigma_a^2}{2}\,\partial_{\theta\theta}\, q_{\varphi}(z,\theta) =   q_{\varphi}(z,\theta) - z\,\Theta(2\pi-\varphi-\theta) [q_{\varphi}(z,\theta)+1]\,,\quad 0\leq \theta < 2\pi\,,\label{eq:dsediffg}
\end{align} 
along with the periodic boundary conditions
\begin{align}
  q_{\varphi}(z,\theta=0) &= q_{\varphi}(z,\theta=2\pi)\,,\nonumber\\
 \partial_{\theta} q_{\varphi}(z,\theta=0) &= \partial_{\theta} q_{\varphi}(z,\theta=2\pi)\,.\label{eq:dsebc}
\end{align}
Due to the presence of the Heaviside step function, the differential equation (\ref{eq:dsediffg}) can be solved separately on the two domains $\theta\in [0,2\pi-\varphi]$ and $\theta \in ]2\pi-\varphi,2\pi[$ and we find the general solution
\begin{align}
  q_{\varphi}(z,\theta) = \left\{\begin{array}{ll}
 \frac{z}{1-z}+A(z,\varphi) \,e^{\frac{\sqrt{2}\sqrt{1-z}\,\theta}{\sigma_a}}+ B(z,\varphi)\, e^{-\frac{\sqrt{2}\sqrt{1-z}\,\theta}{\sigma_a}} \,,& 0\leq\theta\leq 2\pi-\varphi\,,\\[1em]
  C(z,\varphi) \,e^{\frac{\sqrt{2}\theta}{\sigma_a}}+ D(z,\varphi)\, e^{-\frac{\sqrt{2}\theta}{\sigma_a}}\,,& 2\pi-\varphi<\theta< 2\pi\,,
  \end{array}\right.\label{eq:dsesol}
\end{align}
where the integration constants $A(z,\varphi)$, $B(z,\varphi)$, $C(z,\varphi)$ and $D(z,\varphi)$ can be found by using the periodic conditions (\ref{eq:dsebc}) along with the continuity of the solution and its derivative at $2\pi-\varphi$. These four conditions give four equations for the four integration constants, which can be summarised in the matrix form
\begin{align}
  \mathbf{\Omega}(z,\varphi)
  \left(
\begin{array}{c}
 A(z,\varphi)\\
 B(z,\varphi)\\
 C(z,\varphi)\\
 D(z,\varphi)
\end{array} 
\right) =   \left(
\begin{array}{c}
 \frac{z}{z-1}\\
 0\\
 \frac{z}{z-1}\\
 0
\end{array} 
\right)\,,\label{eq:sysO}
\end{align}
where the matrix $\mathbf{\Omega}(z,\varphi)$ is given by
\begin{align}
\mathbf{\Omega}(z,\varphi)=  \left(
\begin{array}{cccc}
 1 & 1 & -e^{\frac{2\sqrt{2}\pi}{\sigma_a}} & -e^{-\frac{2\sqrt{2}\pi}{\sigma_a}} \\
 \sqrt{1-z} & -\sqrt{1-z} &-\,e^{\frac{2\sqrt{2}\pi}{\sigma_a}} & e^{-\frac{2\sqrt{2}\pi}{\sigma_a}} \\
 e^{\frac{\sqrt{2}\sqrt{1-z}(2\pi-\varphi)}{\sigma_a}} & e^{-\frac{\sqrt{2}\sqrt{1-z}(2\pi-\varphi)}{\sigma_a}} & -e^{\frac{\sqrt{2}(2\pi-\varphi)}{\sigma_a}} &
- e^{-\frac{\sqrt{2}(2\pi-\varphi)}{\sigma_a}}   \\
\sqrt{1-z}\,e^{\frac{\sqrt{2}\sqrt{1-z}(2\pi-\varphi)}{\sigma_a}}& -\sqrt{1-z}\,e^{\frac{-\sqrt{2}(2\pi-\varphi)}{\sigma_a}}  & -\,e^{\frac{\sqrt{2}(2\pi-\varphi)}{\sigma_a}}
    & e^{-\frac{\sqrt{2}(2\pi-\varphi)}{\sigma_a}}\\
\end{array} 
\right)\,.\label{eq:Omega}
\end{align}
This set of linear equations can be solved and yields exact expressions that are too long to be displayed. In the next section, we will obtain the mean first-passage time and analyse the linear system (\ref{eq:sysO}) to obtain the behavior of the survival probability in the limit of a large number of steps $n$.

\subsubsection{Mean first-passage time}

\begin{figure}[t]
  \begin{center}
\subfloat{%
 \includegraphics[width=0.5\textwidth]{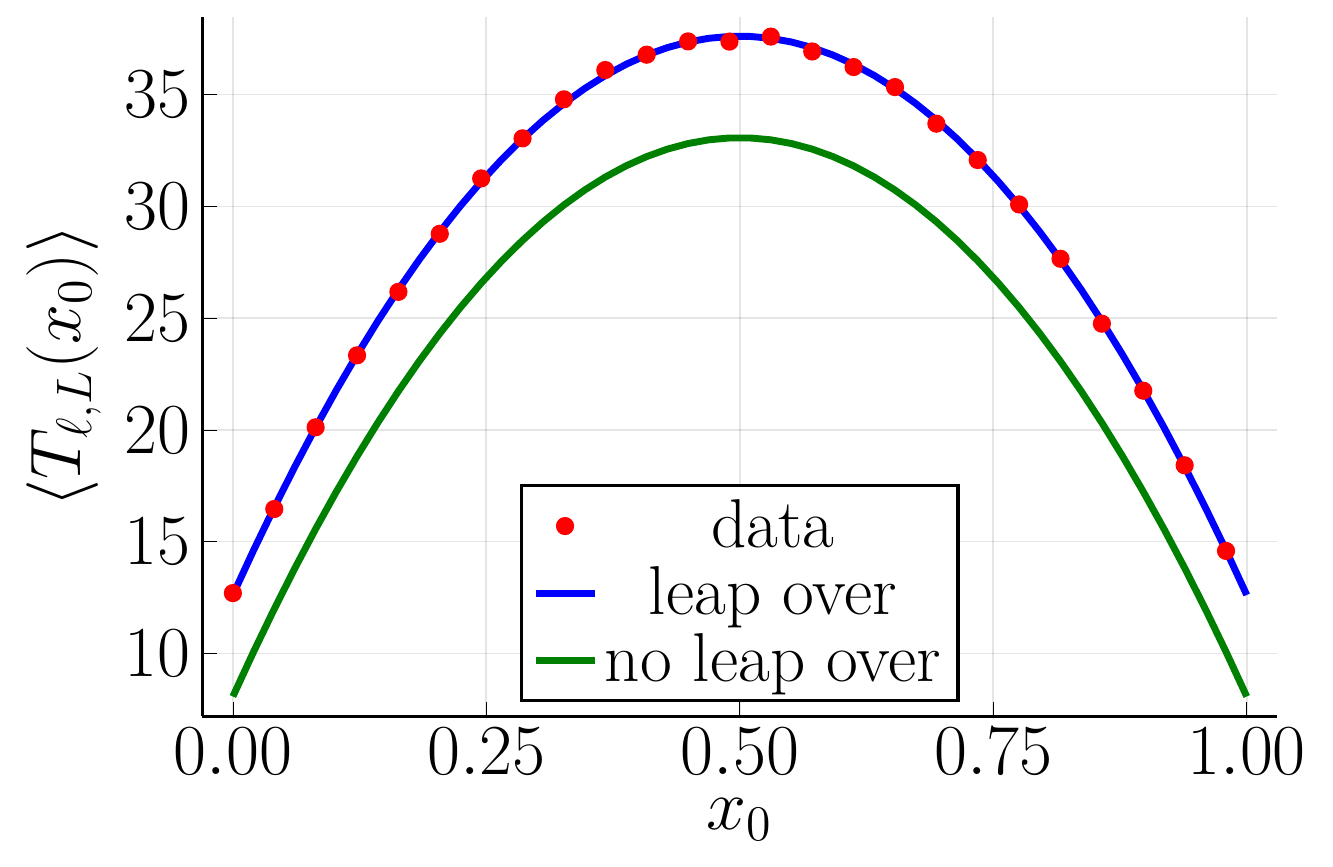}%
}\hfill
\subfloat{%
  \includegraphics[width=0.5\textwidth]{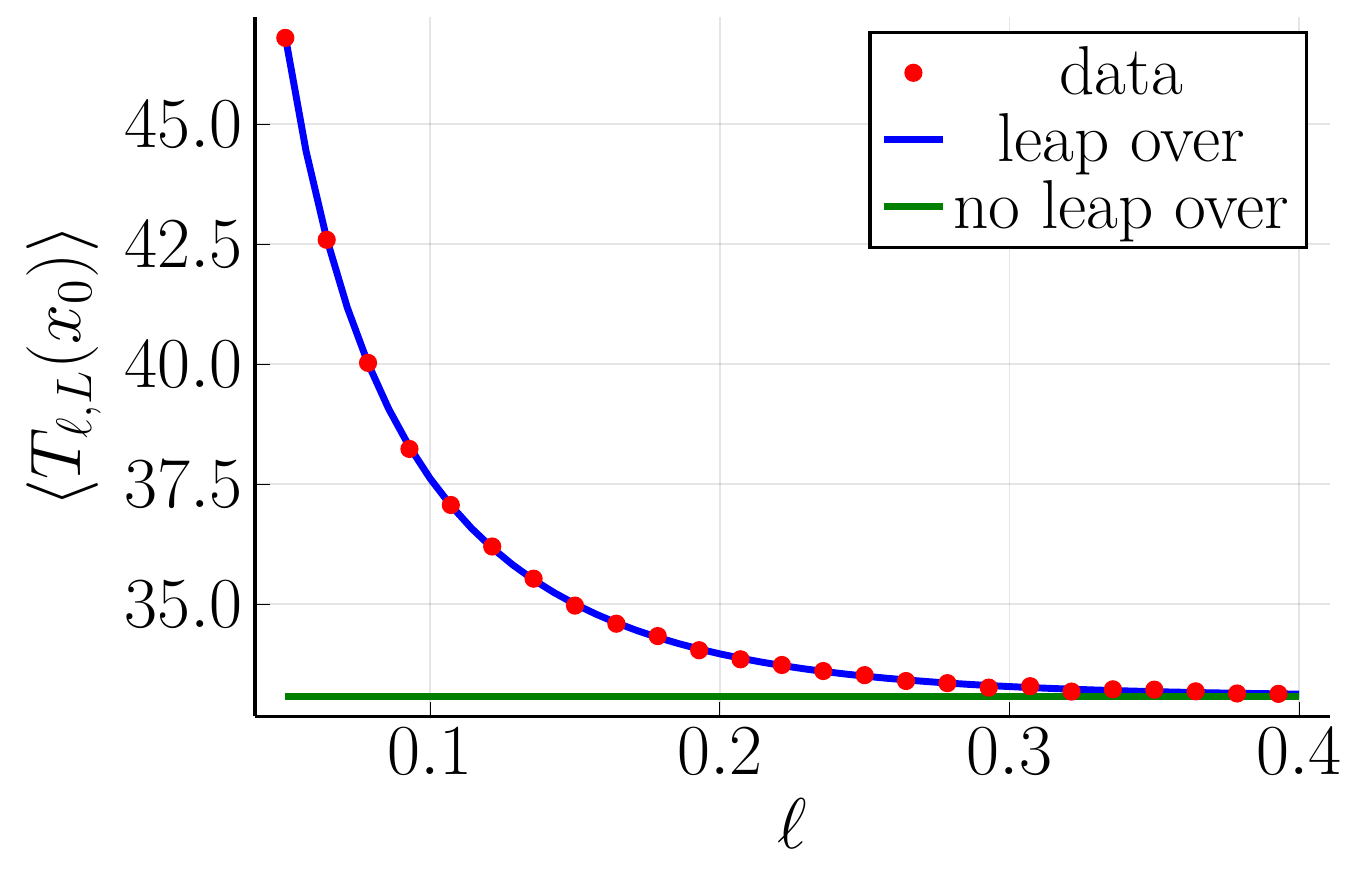}%
}\hfill
    \caption{ Mean first-passage time $\langle T_{\ell,L}(x_0)\rangle $ of a random walk leaping over periodically distributed traps of length $\ell$ given that it started from $x_0$. In the left panel, the mean first-passage time is shown as a function of the initial position $x_0$ with a fixed trap length $\ell=10^{-1}$. In the right panel, the mean first-passage time is shown as a function of the trap length $\ell$ with a fixed initial position $x_0=0.5$. In both panels, the mean first-passage time obtained numerically (red dots) is compared with the theoretical prediction (blue line) given in (\ref{eq:mpftsolx}). As a reference, the mean first-passage time for a random walk without leap overs (green line) given in (\ref{eq:mpftsolx_nojumps}) is displayed. The jump distribution is the double-sided exponential distribution (\ref{eq:dse}) with $\sigma=10^{-1}$ and the distance between the traps is set to $L=1$. }
    \label{fig:MFPT}
  \end{center}
\end{figure}

The mean first-passage time $\langle T_{\varphi}(\theta)\rangle$ for the random walk to be absorbed by the trap of angular size $\varphi$, given that it started at an angular position $\theta$ in the periodic environment given in figure \ref{fig:circle}, is obtained by summing the survival probability (\ref{eq:Sc}) over $n$ \cite{rednerGuide,bray2013persistence}
\begin{align}
  \langle T_{\varphi}(\theta)\rangle =  \sum_{n=0}^\infty S_{\varphi}(n\,|\,\theta) = q_{\varphi}(1,\theta) + 1\,,\label{eq:mfpt}
\end{align}
where we used the definition of the generating function in (\ref{eq:dsegen}) and where the additional $1$ comes from the fact that the sum in the generating function starts from $n=1$. In principle, the generating function $q_{\varphi}(1,\theta)$ can be obtained by setting $z=1$ in its expression given in (\ref{eq:dsesol}). This turns out to be a subtle calculation which requires a careful analysis of the linear system (\ref{eq:sysO}) in the vicinity of $z=1$ (see \ref{app:mfpt}). Alternatively, $q_{\varphi}(1,\theta)$ can easily be obtained by solving the differential equation (\ref{eq:dsediffg}) for $z=1$. Here, we follow the latter approach and replace $q_{\varphi}(z=1,\theta)$ with $\langle T_{\varphi}(\theta)\rangle$ in the differential equation (\ref{eq:dsediffg}), which gives
\begin{align}
  \frac{\sigma_a^2}{2}\partial_{\theta\theta}\,\langle T_{\varphi}(\theta)\rangle =   \langle T_{\varphi}(\theta)\rangle-1 - \,\Theta(2\pi-\varphi-\theta) \langle T_{\varphi}(\theta)\rangle\,,\quad 0\leq \theta < 2\pi\,,\label{eq:diffq1}
\end{align}
along with the periodic boundary conditions (\ref{eq:dsebc}) evaluated at $z=1$
\begin{align}
  \langle T_{\varphi}(\theta=0)\rangle &= \langle T_{\varphi}(\theta=2\pi)\rangle\,,\nonumber\\
 \partial_{\theta} \langle T_{\varphi}(\theta=0)\rangle &= \partial_{\theta} \langle T_{\varphi}(\theta=2\pi)\rangle\,.\label{eq:dsebc1}
\end{align}
The differential equation (\ref{eq:diffq1}) can be solved separately on the two domains $\theta\in [0,2\pi-\varphi]$ and $\theta \in ]2\pi-\varphi,2\pi[$ and we find the general solution
\begin{align}
  \langle T_{\varphi}(\theta)\rangle = \left\{\begin{array}{ll}
 -\frac{\theta^2}{\sigma_a^2}+A(\varphi)\, \frac{\theta}{\sqrt{2}\sigma_a} + B(\varphi)\,,& 0\leq\theta\leq 2\pi-\varphi\,,\\[1em]
 C(\varphi) \,e^{\frac{\sqrt{2}\theta}{\sigma_a}}+ D(\varphi)\, e^{-\frac{\sqrt{2}\theta}{\sigma_a}}+1\,,& 2\pi-\varphi<\theta<2\pi\,,
  \end{array}\right.\label{eq:sol1}
\end{align}
where the integration constants can be found by using the periodic conditions (\ref{eq:dsebc1}) along with the continuity of the solution and its derivative at $2\pi-\varphi$. These equations give the linear system
\begin{align}
 \left(
\begin{array}{cccc}
 0 & 1 & -e^{\frac{2\sqrt{2}\pi}{\sigma_a}} & -e^{-\frac{2\sqrt{2}\pi}{\sigma_a}} \\
1 & 0 &-2e^{\frac{2\sqrt{2}\pi}{\sigma_a}} & 2e^{-\frac{2\sqrt{2}\pi}{\sigma_a}} \\
\frac{2\pi-\varphi}{\sqrt{2}\sigma_a}& 1 & -e^{\frac{\sqrt{2}(2\pi-\varphi)}{\sigma_a}}&-e^{-\frac{\sqrt{2}(2\pi-\varphi)}{\sigma_a}}   \\
1& 0 & -2e^{\frac{\sqrt{2}(2\pi-\varphi)}{\sigma_a}}& 2e^{-\frac{\sqrt{2}(2\pi-\varphi)}{\sigma_a}}\\
\end{array} 
\right)
\left(
\begin{array}{c}
 A(\varphi)\\
 B(\varphi)\\
 C(\varphi)\\
 D(\varphi)
\end{array} 
\right) =   \left(
\begin{array}{c}
 1\\
0\\
 \frac{(2\pi-\varphi)^2}{\sigma_a^2}+1\\
\frac{2\sqrt{2}(2\pi-\varphi)}{\sigma_a^2}
\end{array} 
\right)\,.\label{eq:linmpft}
\end{align}
By solving the linear system (\ref{eq:linmpft}),  we find that the mean first-passage time is given by
\begin{align}
  \langle T_{\varphi}(\theta)\rangle = 1  + \frac{\theta(2\pi-\varphi-\theta)}{\sigma_a^2}+\frac{(2\pi-\varphi)}{\sqrt{2}\sigma_a}\,\coth\left(\frac{\varphi}{\sqrt{2}\sigma_a}\right)\,,\quad 0\leq \theta\leq 2\pi-\varphi\,, \label{eq:mpftsol}
\end{align}
where we restricted the solution to the physically relevant domain $\theta \in [0, 2\pi-\varphi]$. In particular, we can check that $\langle T_{\varphi}(0)\rangle=\langle T_{\varphi}(2\pi-\varphi)\rangle$, which is enforced by symmetry. Going back to the original coordinates of the problem $x_0$, $\ell$, $\sigma$ and $L$, this gives 
 \begin{align}
  \langle T_{\ell,L}(x_0)\rangle &= 1  + \frac{x_0(L-x_0)}{\sigma^2}+\frac{L}{\sqrt{2}\sigma}\,\coth\left(\frac{\ell}{\sqrt{2}\sigma}\right)\,,\quad& 0 \leq x_0 \leq L\,, \label{eq:mpftsolx}
\end{align}
where we used the relations (\ref{eq:varphi}), (\ref{eq:theta0}) and (\ref{eq:sigmaa}). This result is in excellent agreement with numerical simulations (see figure \ref{fig:MFPT}).

In order to appreciate the physical significance of the result (\ref{eq:mpftsolx}), it is useful to compare it with the mean first-passage time of the random walk if it was not allowed to leap over the traps. 
 Note that in this case, the mean first-passage time $\langle T_L(x_0)\rangle_{\text{no leap over}}$ is simply the time it takes to exit an interval of length $L$, given that it started at $0\leq x_0\leq L$. In principle, this time can be obtained by solving a similar differential equation as we did above (see \ref{app:nojump}). Alternatively, one can simply take the limit $\ell\rightarrow\infty$ in the mean first-passage time (\ref{eq:mpftsolx}), which gives
\begin{align}
   \langle T_L(x_0)\rangle_{\text{no leap over}}&= 1  + \frac{x_0(L-x_0)}{\sigma^2}+\frac{L}{\sqrt{2}\sigma}\,\,,\quad &0\leq x_0 \leq L\,,\label{eq:mpftsolx_nojumps}
\end{align}
which coincides with equation (21) in \cite{Masoliver(2005)}.  
Hence we see that the ability of the random walk to leap over traps increases the mean first-passage time by the hyperbolic cotangent term in (\ref{eq:mpftsolx}), which interestingly does not depend on the initial position $x_0$. This term diverges as the trap length goes to $\ell \rightarrow 0$ such that the mean first-passage time behaves as $\langle T_{\ell,L}(x_0)\rangle\sim L/\ell$ in the limit of small traps. Interestingly, similar results have been observed in the context of fully chaotic dynamical systems. In \cite{Nagler(2007)}, the authors consider a different but related problem where they study the mean first-passage time of ballistic particles in a Bunimovich stadium billiard, i.e.~a rectangle billiard capped by two semicircular arcs with reflective boundaries, equipped with a circular hole of radius $\epsilon$. By averaging over the initial phase space configurations, they obtain a mean first-passage time for the particle to fall into the hole. Interestingly, when the hole is much smaller than the system size and independently of its position, the mean first-passage time behaves asymptotically as $1/\epsilon$, which is the same behavior as in our problem upon identifying $\epsilon$ with $\ell$. 

\subsection{Tail of the survival probability}
\label{sec:tail}
Due to the presence of the traps, we expect the survival probability to decay exponentially for large $n$ and we wish to determine the rate $\alpha(\ell,L)$ defined by
\begin{align}
 \alpha(\ell,L) \coloneqq \lim_{n\rightarrow \infty} -\frac{\ln S_{\ell,L}(n\,|\,x_0) }{n} \,.\label{eq:Sdecay}
\end{align}
This rate can be obtained by inspecting the behavior of the generating function (\ref{eq:dsesol}) for $z>1$. Indeed, as we expect the survival probability to decay like $S_{\ell,L}(n\,|\,x_0) \propto e^{-\alpha(\ell,L)\,n}$, the generating function will diverge as $\frac{1}{z-e^{\alpha(\ell,L)}}$ for $z\rightarrow e^{\alpha(\ell,L)}>1$.  It is clear from the expression of the generating function (\ref{eq:dsesol}) that this divergence can only come from the integration constants, which means that the linear system (\ref{eq:sysO}) becomes ill-defined for $z\rightarrow e^{\alpha(\ell,L)}$. Therefore, $e^{\alpha(\ell,L)}$ can be identified as a zero of the determinant of the matrix $\mathbf{\Omega}(z,\varphi)$ (\ref{eq:Omega}), which is given by
\begin{align}
  \text{det}[\mathbf{\Omega}(z,\varphi)] &= 4 (z-2) \sinh\left(\frac{\sqrt{2}\varphi}{\sigma_a}\right) \sinh \left(\frac{\sqrt{2}\,(2\pi-\varphi) \sqrt{1-z}}{\sigma_a}\right)\nonumber \\
  &\quad +8 \sqrt{1-z}\left[1-\cosh\left(\frac{\sqrt{2}\varphi}{\sigma_a}\right) \cosh \left(\frac{\sqrt{2}\,(2\pi-\varphi) \sqrt{1-z}}{\sigma_a}\right)\right]\,.\label{eq:det}
\end{align}
Setting the determinant to zero and going back to the initial coordinates $\ell$, $\sigma$ and $L$, we find that the decay rate satisfies the transcendental equation
\begin{align}
  &(e^{\alpha(\ell,L)}-2)\sinh\left(\frac{\sqrt{2}\ell}{\sigma}\right) \sin \left(\frac{\sqrt{2}\,L\sqrt{e^{\alpha(\ell,L)}-1}}{\sigma}\right)\nonumber \\
  &\quad +2 \sqrt{e^{\alpha(\ell,L)}-1}\left[1-\cosh\left(\frac{\sqrt{2}\ell}{\sigma}\right) \cos \left(\frac{\sqrt{2}\,L \sqrt{e^{\alpha(\ell,L)}-1}}{\sigma}\right)\right] = 0\,,\label{eq:detzero}
\end{align}
where we used the relations (\ref{eq:varphi}), (\ref{eq:theta0}) and (\ref{eq:sigmaa}).
The equation (\ref{eq:detzero}) has a multitude of zeros, which corresponds to the different modes of the survival probability. The decay rate $\alpha(\ell,L)$ corresponds to the lowest zero that is strictly positive. It seems difficult to find an analytical expression of this zero for arbitrary values of the trap length $\ell$ and distance $L$. Nevertheless, we can make progress in the small and large traps limits, while keeping $L$ fixed. We present these two limits in the remaining of this section.

\paragraph{Small traps limit $\ell \rightarrow 0$}  In this limit, we expect the survival probability to approach $1$ as the trap length becomes vanishingly small. Therefore, we can look for a perturbative expansion of the decay rate of the form 
\begin{align}
  \alpha(\ell,L)\sim a_1(L)\,\ell+a_2(L)\,\ell^2+O(\ell^3)\,,\quad \ell \rightarrow 0\,,\label{eq:expa}
\end{align}
where $a_1(L)$ and $a_2(L)$ are coefficients, independent of $\ell$, that remain to be determined. 
 By inserting the perturbative expansion in the transcendental equation, we find, at first order, an equation satisfied by $a_1(L)$
\begin{align}
a_1(L) L-1 = 0\,,\label{eq:forder}
\end{align}
which gives $a_1(L)=1/L$.  At second order, the transcendental equation provides an equation satisfied by $a_2(L)$ which reads
\begin{align}
 3 \sigma ^2 \left(2 a_2(L)\, L^2+1\right)+L^2=0\,,\label{eq/sorder}
\end{align}
which gives $a_2(L)=-\frac{L^2+3 \sigma ^2}{6 L^2 \sigma ^2}$. Inserting the values of the coefficients $a_1(L)$ and $a_2(L)$ in the perturbative expansion of the decay rate, we find
\begin{align}
\alpha(\ell,L) \sim \frac{\ell}{L}-\left(\frac{1}{2L^2} + \frac{1}{6\sigma^2}\right)\ell^2\,, \quad \ell \to 0\,.\label{eq:Salg}
\end{align}
Similarly to the mean first-passage time, the leading order term in this expansion is the same as the one that is usually observed in ``open" dynamical systems such as classical billiards in $2d$ with reflecting boundaries \cite{Dettmann(2011)}. In such systems, particles travel at constant speed, save at collisions, until they hit an absorbing hole. Upon averaging over uniform  initial phase space configurations, it is possible to define a survival probability, which also decays exponentially with a rate that depends on the hole size and the perimeter of the billiard. In the limit of small holes, the decay rate tends to the ratio of the size of the hole over the size of the billiard \cite{Nagler(2007), Bunimovich(2007)}, which can be identified to the leading order ratio $\ell/L$ in our stochastic model. At the second order, the geometrical term $-\frac{\ell^2}{2L^2}$ is also present in the decay rate of the classical billiard. The additional term $- \frac{\ell^2}{6\sigma^2}$ in the second order is related to the dynamics and corresponds, in the deterministic context, to an infinite series of correlations due to the map describing the time evolution of collisions \cite{Bunimovich(2007)}.

\begin{figure}[t]
  \begin{center}
    \includegraphics[width=0.5\textwidth]{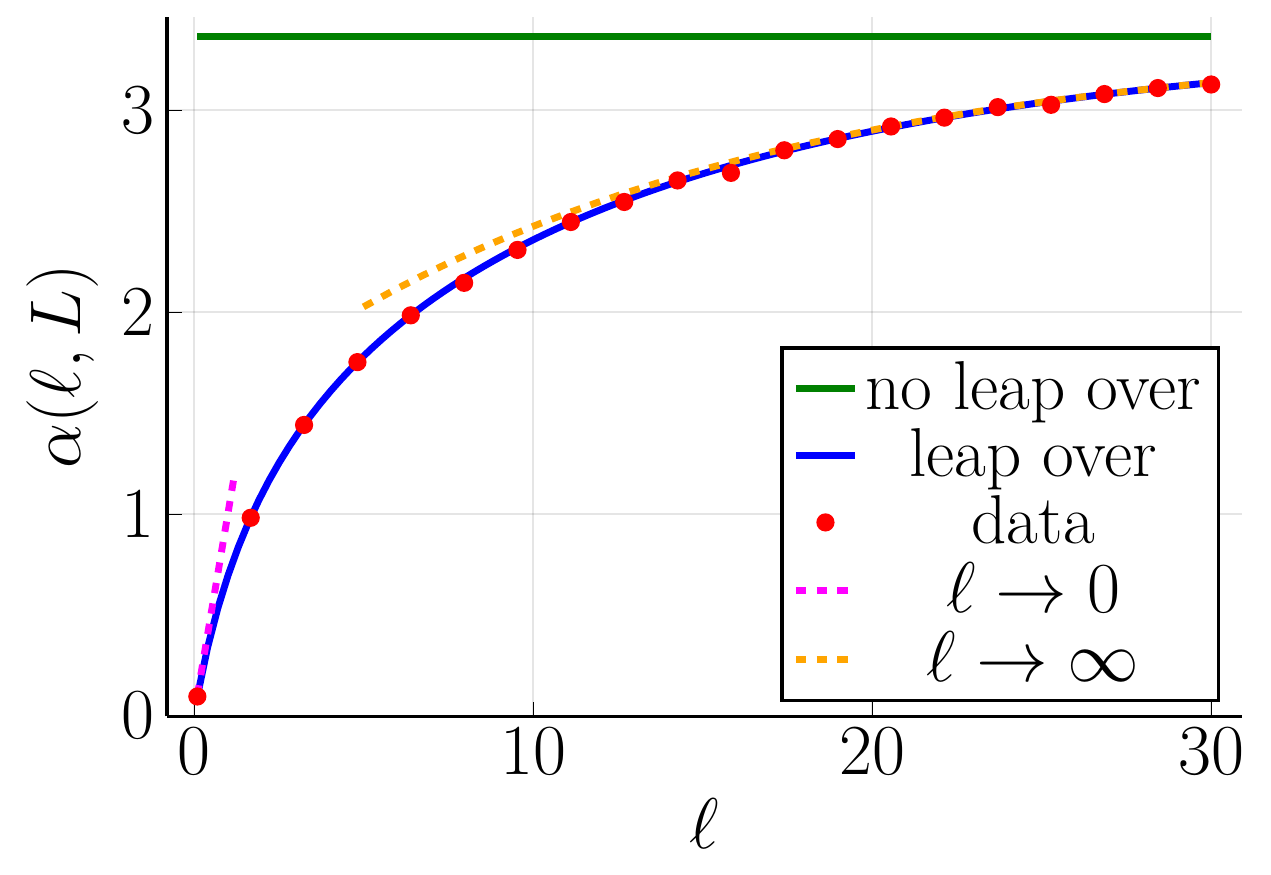}
    \caption{Decay rate $\alpha(\ell,L) =\lim_{n\rightarrow \infty} -\frac{\ln S_{\ell,L}(n\,|\,x_0) }n$ of the survival probability $S_{\ell,L}(n\,|\,x_0)$ for a large number of steps $n$. The numerical decay rate (red dots) is compared to the theoretical decay rate (blue line), satisfying the transcendental equation (\ref{eq:detzero}), as a function of the trap length $\ell$ for $L=1$ and $\sigma=20$. The theoretical asymptotic behaviors for $\ell\rightarrow 0$, given in (\ref{eq:Salg}), and $\ell \rightarrow \infty$, given in (\ref{eq:Sexpd2}), are also represented (dashed lines). As a reference the decay rate for a random walk without leap overs (green line), satisfying the transcendental equation (\ref{eq:det2pi}), is displayed.}
    \label{fig:alpha}
  \end{center}
\end{figure}

\paragraph{Large traps limit $\ell \rightarrow \infty$} In this limit we expect the decay rate to converge to a constant $\alpha(\ell,L)\to \alpha^*(L)$ independent of $\ell$ as the length of the traps becomes infinitely large. This constant $\alpha^*(L)$ is the solution of the transcendental equation (\ref{eq:detzero}) in the limit of $\ell \rightarrow \infty$:
\begin{align}
 (e^{\alpha^*(L)}-2) \sin \left(\frac{\sqrt{2}\,L\sqrt{e^{\alpha^*(L)}-1}}{\sigma}\right) -2 \sqrt{e^{\alpha^*(L)}-1} \cos \left(\frac{\sqrt{2}\,L \sqrt{e^{\alpha^*(L)}-1}}{\sigma}\right)= 0\,,\label{eq:det2pi}
\end{align}
which is independent of $\ell$, as expected. This decay rate corresponds to the random walk which is not allowed to leap over traps, i.e. a random walk on an finite interval of length $L$ with absorbing boundaries. We can further examine the transcendental equation in the limit of $\ell\rightarrow \infty$ to obtain a second order correction which reads
\begin{align}
  \alpha(\ell,L) \sim \alpha^*(L)-\frac{8 \,e^{-2 \alpha^*(L) } \left(2-e^{\alpha^*(L) }\right) \left(1-e^{\alpha^*(L) }\right) }{\left(\frac{\sqrt{2} Le^{\alpha^*(L) }}{\sigma} +2\right) \cos \left(\frac{\sqrt{2} L\sqrt{e^{\alpha^*(L) }-1} }{\sigma }\right) }\,e^{-\frac{\sqrt{2}\ell}{\sigma}}\,,\qquad \ell\to \infty\,,\label{eq:Sexpd2}
\end{align}
where $\alpha^*(L)$ is the solution of the transcendental equation (\ref{eq:det2pi}). The exact expression for the decay rate and its asymptotic behaviors for $\ell\rightarrow 0$  and $\ell \rightarrow \infty$ are in excellent agreement with numerical simulations (see figure \ref{fig:alpha}).

Note that it is also possible to derive the asymptotic behaviors of the decay rate $\alpha(\ell,L)$ for $L\to 0$ and $L\to \infty$ with $\ell$ fixed (see \ref{app:L0I}). The asymptotic behaviors are given by
\begin{align}
  \alpha(\ell,L) &\sim \ln\left(\frac{1}{L}\right)\,,\qquad L\to 0\,,\label{eq:alpha0m}\\
   \alpha(\ell,L) &\sim \frac{\pi^2\sigma^2}{2L^2}\,,\qquad L\to \infty\,.\label{eq:alphaLim}
\end{align}
In the case of $L\to 0$, the survival probability decays as $S_{\ell,L}(n\,|\,x_0)\propto L^n$ when $L\rightarrow 0$, which is expected as the traps become infinitely close to each others. In the case of $L\to \infty$, the fact that the decay rate tends to zero while a trap is still present close to the origin signifies a commutation issue with the limits $L\rightarrow \infty$ and $n\rightarrow\infty$ (in the definition of the decay rate (\ref{eq:Sdecay})). Indeed, if we perform the limit $L\rightarrow \infty$ first, one would expect a power-law decay of the survival probability in the limit of $n \to \infty$. From the expression (\ref{eq:alphaLim}), we can infer that there should be a scaling function with a scaling variable $n\,\sigma^2/L^2$ that interpolates between the two ways of ordering the limits.

\section{Continuous-time random walk model (CTRW)}\label{sec:ctrw}
In this section, we extend our results to the case of CTRWs.
We introduce a waiting period before each jump that follows a heavy-tailed distribution 
\begin{align}
  \psi(t)\sim b \,t^{-1-\beta}\,,\quad t\to \infty\,,\label{eq:psit}
\end{align} 
with $\beta>0$ and $b>0$. As in the previous section, the jump distribution is the double-sided exponential distribution given in (\ref{eq:dse}). We study the survival probability and the mean first-passage time of this CTRW. We first show that the heavy-tailed waiting time distribution induces an algebraically decay of the survival probability. Then, we discuss the existence of the mean first-passage time and provide its expression. 

\subsection{Survival probability}
 The survival probability $S_{\ell,L}(t\,|\,x_0)$ after a time $t$ can be obtained by summing over all possible number of steps $n$ weighted by the probability distribution $\chi_n(t)$ that exactly $n$ steps were made during the time $t$: 
\begin{align}
S_{\ell,L}(t\,|\,x_0)=\sum_{n=0}^{\infty}S_{\ell,L}(n\,|\,x_0) \chi_n(t)\,.\label{eq:SCTRW}
\end{align}
The probability distribution $\chi_n(t)$ that exactly $n$ steps were made up to the time $t$ can be obtained as follows. For the random walk to make $n$ steps up to time $t$, it must first make the $n^\text{th}$ step at an earlier time $t'<t$ and then remain at the same position for the remaining time $t-t'$, which happens with probability $1-\int_0^{t-t'}d\tau\,\psi(\tau)$. Denoting $\psi_n(t')$ the probability distribution that the $n^\text{th}$ step is made at the time $t'$, it reads
\begin{align}
\chi_n(t)=\int_0^tdt'\psi_n(t')\left[1-\int_0^{t-t'}d\tau\,\psi(\tau)\right]\,.\label{eq:chi_n(t)}
\end{align}
To obtain the probability distribution $\psi_n(t')$ of the time $t'$ at which the $n^\text{th}$ step is made, we use the following recurrence relation, which states that the random walk must first make the $(n-1)^\text{th}$ step at an earlier time $\tau$ and then make the $n^\text{th}$ step after the remaining time $t'-\tau$:
\begin{align}
\psi_n(t')= \int_0^{t'} d\tau\,\psi_{n-1}(\tau)\psi(t'-\tau)\,.\label{eq:psi_n(t)}
\end{align}
The recurrence relation (\ref{eq:psi_n(t)}) can be solved in Laplace domain and gives
\begin{align}
  \tilde \psi_n(s) = \int_0^\infty dt' e^{-st'}  \psi_n(t') = \tilde \psi^n(s)\,,\label{eq:solpsin}
\end{align}
where $\tilde\psi(s)=\int_0^\infty dt e^{-st}\psi(t)$ is the Laplace transform of the waiting time distribution. The Laplace transform of the distribution $\chi_n(t)$ can now be obtained by taking the Laplace transform of (\ref{eq:chi_n(t)}) and inserting the expression of $\tilde \psi_n(s)$ obtained in (\ref{eq:solpsin}), which gives
\begin{align}
 \tilde \chi_n(s) = \psi^n(s)\,\frac{1-\tilde \psi(s)}{s}\,,\label{eq:chissol}
\end{align}
where we used that a convolution becomes a product in Laplace domain.
Taking a Laplace transform of the survival probability (\ref{eq:SCTRW}) and inserting the expression of $\tilde \chi_n(s)$ obtained in (\ref{eq:chissol}), we find an exact expression for the Laplace transform of the survival probability:
\begin{align}
  \tilde S_{\ell,L}(s\,|\,x_0) = \frac{1-\tilde \psi(s)}{s}\sum_{n=0}^{\infty}S_{\ell,L}(n\,|\,x_0)\, \tilde\psi^n(s)\,.\label{eq:SCTRWs2}
\end{align}
To investigate the long time limit of $S_{\ell,L}(t\,|\,x_0)$, we need to consider the small $s$ limit of its Laplace transform. In this limit, we expect the sum in (\ref{eq:SCTRWs2}) to be dominated for large $n$. Therefore, we replace the survival probability in (\ref{eq:SCTRWs2}) by its large $n$ behavior, i.e. an exponential decay $S_{\ell,L}(n\,|\,x_0)\propto e^{-n\,\alpha(\ell,L)} $ where the decay rate $\alpha(\ell,L)$ was computed in the section \ref{sec:tail}. This gives a geometric sum which yields
\begin{align}
   \tilde S_{\ell,L}(s\,|\,x_0) &\propto \frac{1-\tilde \psi(s)}{s}\sum_{n=0}^{\infty} e^{-n\alpha(\ell,L)}\, \tilde\psi^n(s)\,,\nonumber\\
   &\propto \frac{1-\tilde \psi(s)}{s\,[1-\tilde \psi(s)\,e^{-\alpha(\ell,L)}]}\,,\qquad s\to 0\,.\label{eq:SCTRWs3}
\end{align}
Note the use of the proportionality sign $\propto$ since we omit an overall coefficient, independent of $n$, when we use the large $n$ expression of the survival probability $S_{\ell,L}(n\,|\,x_0)\propto e^{-n\,\alpha(\ell,L)}$. In principle, it is possible to obtain the overall coefficient by solving the survival probability exactly in Laplace domain (see \ref{app:LapCTRW}). Since we are only interested in the decay behavior of the survival probability, we omit the overall constant and investigate the survival probability for $t\rightarrow \infty$, which corresponds to $s\rightarrow 0$ in the Laplace transform (\ref{eq:SCTRWs3}). Since the waiting time distribution behaves as $\psi(t)\sim b \,t^{-1-\beta}$ for large $t$, we have that the non-analytical part of its Laplace transform behaves as $s^\beta$ for $s\rightarrow 0$. Consequently, by inserting this result in (\ref{eq:SCTRWs3}) and by applying Tauberian theorem, it tells us that the survival probability decays algebraically for $t\rightarrow\infty$ as
\begin{align}
    S_{\ell,L}(t\,|\,x_0) &\propto t^{-\beta}\,,\qquad t\to\infty\,.\label{eq:SCTRWs5}
\end{align}
Note that this expression is true for all $\beta>0$. In particular, in the case when $\beta$ is an integer, the small $s$ behavior of the Laplace transform of the waiting time distribution remains non-analytic as it will contain terms of the form $\ln(s)\,s^{\beta}$, which ensure that the result (\ref{eq:SCTRWs5}) remains valid for all $\beta>0$. We observe that the exponent in (\ref{eq:SCTRWs5}) is independent of the trap length $\ell$ and their distances $L$. The signature of $\ell$ and $L$ can probably be found in the overall coefficient.  

\subsection{Mean first-passage time}
To obtain the mean first-passage time, we can in principle extend the differential equations obtained in section \ref{sec:surv} to the CTRW formalism \cite{Masoliver(2005),Montero(2007)} (see \ref{app:LapCTRW}). Alternatively, we can perform a direct computation based on the discrete-time random walk in the following way. The first-passage being an arrival event, the mean first-passage time $\langle T_{\ell,L}(x_0)\rangle_{\text{CTRW}}$ can be obtained by summing the first-passage distribution $f_{\ell,L}(n\,|\,x_0)$ of the $n^\text{th}$ step with the time distribution $\psi_n(t)$  and averaging over time, which gives
\begin{align}
  \langle T_{\ell,L}(x_0)\rangle_{\text{CTRW}} &= \int_0^{\infty} dt\, t \sum_{n=1}^{\infty} f_{\ell,L}(n\,|\, x_0)\psi_n(t)\,, \label{eq:Tctrw}
\end{align}
where the first-passage distribution $f_{\ell,L}(n\,|\, x_0)$ is related to the survival probability discussed in section \ref{sec:surv} through the relation $f_{\ell,L}(n\,|\,x_0)=S_{\ell,L}(n-1\,|\,x_0)-S_{\ell,L}(n\,|\,x_0)$. By switching the order of the sum and the integral in (\ref{eq:Tctrw}), we obtain
\begin{align}
\langle T_{\ell,L}(x_0)\rangle_{\text{CTRW}} &=\sum_{n=1}^{\infty} f_{\ell,L}(n\,|\, x_0)\int_0^{\infty} dt\, t \,\psi_n(t)\,,\nonumber\\
&= \langle \tau \rangle\sum_{n=1}^{\infty} n f_{\ell,L}(n\,|\, x_0) =  \langle \tau \rangle\langle T_{\ell,L}(x_0)\rangle\,,\label{eq:mfpt_cont}
\end{align}
where we used that $\int_0^{\infty} dt\, t \,\psi_n(t)=n\, \langle\tau\rangle$ where $\langle\tau\rangle$ is the mean waiting time before each jump $\langle\tau\rangle=\int_0^{\infty} dt\, \psi(t)$, and where we recognised that the last sum is simply the mean first-passage time $\langle T_{\ell,L}(x_0)\rangle$ of the discrete-time model. The mean waiting time $\langle\tau\rangle$ for the heavy-tailed distribution (\ref{eq:psit}) is finite when $\beta>1$ and infinite when $\beta\leq 1$. Therefore, by using the expression of $\langle T_{\ell,L}(x_0)\rangle$ in (\ref{eq:mpftsol}), the mean first-passage time for this CTRW is given for $\beta>1$ by
\begin{align}
  \langle T_{\ell,L}(x_0)\rangle_{\text{CTRW}} = \langle\tau\rangle  + \frac{ \langle\tau\rangle(L-x_0)x_0}{\sigma^2}+\frac{ \langle\tau\rangle L}{\sqrt{2}\,\sigma}\,\coth\left(\frac{\ell}{\sqrt{2}\,\sigma}\right)\,,\quad 0\leq x_0\leq L\,, \label{eq:mpftsolgen}
\end{align}
 and $\langle T_{\ell,L}(x_0)\rangle_{\text{CTRW}} = +\infty$ when $0<\beta\leq 1$. For the mean first-passage time, the continuous-time extension is merely a change of scale with respect to the number of steps when the mean waiting time $\langle\tau\rangle$ is finite.

\section{Diffusive limits of the discrete-time model}\label{sec:dl}
In this section, we derive the diffusive limit of the discrete-time random walk model presented in section \ref{sec:dtrw} in the assumption that the jump distribution $f(\eta)$ is symmetric and has a finite variance $\sigma^2 = \int_{-\infty}^\infty d\eta\,f(\eta)\,\eta^2$. To do so, we denote $\tau$ the time taken to perform a single jump and we consider the usual diffusive limit $\sigma \rightarrow 0$ with $\tau\rightarrow 0$ while keeping the ratio $\sigma^2/\tau$ fixed. We show that two diffusive limits are possible depending on the scaling chosen for the length of the traps:
\begin{enumerate}
  \item letting the length of the traps $\ell$ tend to zero while keeping $\ell/\tau$ fixed and maintaining the distance $L$ between the traps finite, which yields to diffusion with periodically distributed point absorbers represented by Dirac delta functions,
  \item letting both the length of the traps $\ell$ and the distance between the traps $L$ tend to zero while keeping the ratio $\ell/(\tau L)$ fixed, which yields to diffusion with uniform absorption on the real line.
\end{enumerate}
We derive the two diffusive limits in the subsequent paragraphs.

\subsection{Diffusion with periodically distributed point absorbers}
In this section, we consider the diffusive limit
\begin{align}
  \sigma\rightarrow 0\,,\quad \tau \rightarrow 0\,,\quad \ell\rightarrow 0\,,\quad \text{with}\quad D\coloneqq \frac{\sigma^2}{2\tau}\,,\quad \beta\coloneqq\frac{\ell}{\tau}\quad \text{fixed},\label{eq:diff1}
\end{align}
where $D$ is the diffusion coefficient and $\beta$ is the absorption rate. Note that in this limit the length of the trap is much smaller than the typical step size as it is of order $\ell=O(\sigma^2)$, for $\sigma\rightarrow 0$. From the Markov rule (\ref{eq:mr}), the backward equation for the survival probability $S_{\ell,L}(n\,|\,x)$ writes in the position space:
\begin{align}
  S_{\ell,L}(n\,|\,x) = \sum_{m=-\infty}^{\infty} \int_{m\,(L+\ell)-x}^{m\,(L+\ell)+L-x} d\eta\,S_{\ell,L}(n-1\,|\,x+\eta) f(\eta)\,, \label{eq:fordx}
\end{align}
where the sum over $m$ arises from the fact that the random walk can jump to any of the intervals in figure \ref{fig:model}. To derive the diffusive limit, we expand the survival probability in the right-hand side to second order in $\eta$ which gives
\begin{align}
  S_{\ell,L}(n\,|\,x)  \sim S_{\ell,L}(n-1\,|\,x)&\left[\sum_{m=-\infty}^{\infty} \int_{m\,(L+\ell)-x}^{m\,(L+\ell)+L-x} d\eta\, f(\eta)\right] \nonumber\\
&+ \frac{1}{2}\,\partial_{xx}S_{\ell,L}(n-1\,|\,x)\left[\sum_{m=-\infty}^{\infty}\int_{m\,(L+\ell)-x}^{m\,(L+\ell)+L-x} d\eta\, \eta^2 f(\eta)\right]\,,\label{eq:fordd}
\end{align}
where we used that the jump distribution is symmetric to cancel the linear term. In the limit of small $\ell$, the first term in brackets becomes
\begin{align}
 \sum_{m=-\infty}^{\infty} \int_{m\,(L+\ell)-x}^{m\,(L+\ell)+L-x} d\eta\, f(\eta) &\sim 1 -\sum_{m=-\infty}^{\infty}\int_{m(L+\ell)-x-\ell}^{m(L+\ell)-x} d\eta\, f(\eta) \,,\nonumber\\
  &\sim  1 -\ell\sum_{m=-\infty}^{\infty} f(m\,L-x)\,,\nonumber\\
  &\sim  1 -\ell\sum_{m=-\infty}^{\infty}\delta(m\,L-x)\,,\quad \ell\to 0\,,\label{eq:approxD}
\end{align}
where we used the normalisation $\int_{-\infty}^{\infty} d\eta\, f(\eta)=1$ in the first line, the fact that $\ell$ is small in the second line, and that the jump distribution tends to a Dirac delta function in the diffusive limit, as the standard deviation $\sigma$ of the distribution $f(\eta)$ is taken to go to zero, in the third line. The term in the second bracket simply tends towards the variance of the jump distribution in the limit of small $\ell$:
\begin{align}
  \sum_{m=-\infty}^{\infty}\int_{m\,(L+\ell)-x}^{m\,(L+\ell)+L-x} d\eta\, \eta^2 f(\eta) \sim \sigma^2\,,\quad \ell\to 0\,,\label{eq:approxD2}
\end{align}
as we can neglect the $\ell$ dependence in the limits of integration. Inserting the expansions of the two terms in brackets (\ref{eq:approxD}) and (\ref{eq:approxD2}), and going from the number of steps $n$ to a continuous time $t$, we find the diffusion equation with periodically distributed point absorbers
\begin{align}
  \partial_t S_{\beta}(t\,|\,\theta) = D \partial_{xx} S_{\beta}(t\,|\,x)  - \beta \sum_{m=-\infty}^{\infty} \delta(x-m\,L)S_{\beta}(t\,|\,x) \,,\label{eq:difftrap}
\end{align}
 where $D$ and $\beta$ are the diffusion coefficient and absorption rate respectively. In the remaining of this section, we derive the survival probability and the mean first-passage time of this diffusive limit.

\subsubsection{Survival probability} 
As in section \ref{sec:surv}, it is convenient to map the periodic structure of the diffusion equation (\ref{eq:difftrap}) to a circular geometry of perimeter $L$. By performing a change of coordinate $\theta=2\pi x/L$ in the diffusion equation (\ref{eq:difftrap}), we find that the survival probability $S_{\beta_a}(t\,|\,\theta)$ of a diffusive particle on the circle starting at $\theta$ in the presence of a trap located at $\theta=0$ satisfies the equation
\begin{align}
  \partial_t S_{\beta_a}(t\,|\,\theta) = D_a\,\partial_{\theta \theta} S_{\beta_a}(t\,|\,\theta) - \beta_a\, \delta(\theta) S_{\beta_a}(t\,|\,\theta) \,,\quad 0\leq\theta<2\pi\,,\label{eq:survdiff}
\end{align}
with the initial condition $S(t=0\,|\,\theta)=1$ for $\theta \in ]0,2\pi[$ and where $D_a$ and $\beta_a$ are the angular diffusion coefficient and absorption rate given by
\begin{align}
  D_a \coloneqq \frac{4\pi^2D}{L^2}\,,\quad   \beta_a \coloneqq \frac{2\pi\beta}{L}\,, \label{eq:Dbetaa}
\end{align}
where the subscript $a$ stands for ``angular'' diffusion coefficient and absorption rate. Note that the dimensions of $D_a$ and $\beta_a$ are both inverse time.
To solve the equation (\ref{eq:survdiff}), it is convenient to consider it in the Laplace domain in which it reads
\begin{align}
  s \,\tilde S_{\beta_a}(s\,|\,\theta) -1 =  D_a\partial_{\theta \theta} \tilde S_{\beta_a}(s\,|\,\theta) - \beta_a \delta(\theta) \,\tilde S_{\beta_a}(s\,|\,\theta) \,,\quad 0\leq\theta<2\pi\,,\label{eq:survdiffl}
\end{align}
where we used that $\int_0^\infty dt e^{-s\,t} \partial_t S_{\beta_a}(t\,|\,\theta) = s \tilde S_{\beta_a}(s\,|\,\theta) -1$ by integration by parts and by using the initial condition $S(t=0\,|\,\theta)=1$.
To make the equation (\ref{eq:survdiffl}) homogeneous, we shift the solution by $\tilde S_{\beta_a}(s\,|\,\theta)=\frac{1}{s}+U_{\beta_a}(s\,|\,\theta)$ which gives
\begin{align}
   s \,U_{\beta_a}(s\,|\,\theta)  =  D_a\partial_{\theta \theta} U_{\beta_a}(s\,|\,\theta) - \beta_a\, \delta(\theta) \left(\frac{1}{s}+U_{\beta_a}(s\,|\,\theta)\right) \,,\quad 0\leq\theta<2\pi\,.\label{eq:survdiffU}
\end{align}
We now solve the equation (\ref{eq:survdiffU}) on the interval $]0,2\pi[$ where the Dirac delta is absent to find the general solution
\begin{align}
  U_{\beta_a}(s\,|\,\theta) =   A(s,\beta_a) e^{-\sqrt{s/D_a}\,\theta} +  B(s,\beta_a) e^{\sqrt{s/D_a}\,\theta} \,.\label{eq:survdiffUG}
\end{align}
The integration constants are found by imposing periodic boundary conditions on the solution 
\begin{subequations}
\begin{align}
  U_{\beta_a}(s\,|\,0) &= U_{\beta_a}(s\,|\,2\pi)\,,
\end{align}
along with the discontinuity of its derivative obtained by integrating the equation (\ref{eq:survdiffU}) on an infinitesimal interval around $\theta=0$:
\begin{align}
 D_a[\partial_\theta\, U_{\beta_a}(s\,|\,0)-\partial_\theta\,U_{\beta_a}(s\,|\,2\pi)] &= \beta_a\,\left(\frac{1}{s}+U_{\beta_a}(s\,|\,0)\right) \,,
\end{align}
\label{eq:survdiffbc}
\end{subequations}
which gives
\begin{subequations}
\begin{align}
  A(s,\beta_a) &=-\frac{\beta_a}{2s\left[2\sqrt{sD_a}\,\sinh(\pi\sqrt{s/D_a})+\beta_a\, \cosh(\pi\sqrt{s/D_a})\right]}\,e^{\pi\sqrt{s/D_a}}\,,\\
   B(s,\beta_a) &= -\frac{\beta_a}{2s\left[2\sqrt{sD_a}\,\sinh(\pi\sqrt{s/D_a})+\beta_a\, \cosh(\pi\sqrt{s/D_a})\right]}\,e^{-\pi\sqrt{s/D_a}}\,.
\end{align}
\label{eq:survdiffIC}
\end{subequations}
Plugging back the integration constants (\ref{eq:survdiffIC}) into the general solution (\ref{eq:survdiffUG}) we find 
\begin{align}
  U_{\beta_a}(s\,|\,\theta) = -\frac{\beta_a\cosh[\sqrt{s/D_a}(\pi-\theta)]}{s\left[2\sqrt{sD_a}\,\sinh(\pi\sqrt{s/D_a})+\beta_a\, \cosh(\pi\sqrt{s/D_a})\right]}\,,\label{eq:solU}
\end{align}
which in terms of the Laplace transform of the survival probability gives
\begin{align}
  \tilde S_{\beta_a}(s\,|\,\theta) = \frac{1}{s}\left(1 -\frac{\beta_a\cosh[\sqrt{s/D_a}(\pi-\theta)]}{2\sqrt{sD_a}\,\sinh(\pi\sqrt{s/D_a})+\beta_a\, \cosh(\pi\sqrt{s/D_a})}\right)\,.\label{eq:solS}
\end{align}
This Laplace transform seems difficult to invert for arbitrary $t$. Nevertheless, we can study the long time limit, similarly to the discrete case in section \ref{sec:tail}, where we expect an exponential decay of the survival probability. We wish to determine the decay rate $\alpha(\beta,L)$ defined by
\begin{align}
 \alpha(\beta,L) \coloneqq \lim_{t\rightarrow \infty} -\frac{\ln S_{\beta,L}(t\,|\,x_0) }{t}\,.\label{eq:Sdecayb}
\end{align}
This rate can be obtained by inspecting the behavior of the survival probability (\ref{eq:solS}) for $s<0$: as we expect the survival probability to decay like $S_{\beta,L}(t\,|\,x_0) \propto e^{-\alpha(\beta,L)\,t}$, the Laplace transform will diverge as $\tilde S_{\beta,L}(s\,|\,x_0)\propto\frac{1}{s+\alpha(\beta,L)}$ for $s\rightarrow - \alpha(\beta,L) <0$. It is clear by examining (\ref{eq:solS}) that $-\alpha(\beta,L)$ can be identified as the first negative zero of the denominator in (\ref{eq:solS}). Setting the denominator to zero and going back to the initial coordinates $x$, $\beta$, $D$ and $L$, we find that $\alpha(\beta,L)$ satisfies the transcendental equation 
\begin{align}
  \cot\left(\frac L 2\sqrt{\frac{\alpha(\beta,L)}{D}}\right)= \frac{2\sqrt{\alpha(\beta,L)D}}{\beta}\,,\label{eq:transc}
\end{align}
where we have used the relations (\ref{eq:Dbetaa}). Note that the equation (\ref{eq:transc}) can also be derived directly from the discrete result as we show in \ref{app:der diff lim}. From the transcendental equation (\ref{eq:transc}), we see that decay rate will take the scaling form
\begin{equation} 
\alpha(\beta,L)=\frac{D}{L^2} \,\mathcal{G}\left(\text{Sh}=\frac{\beta L}{D}\right)\,,\label{eq:alphascal}
\end{equation}
where $\text{Sh}$ is the Sherwood number and the scaling function $\mathcal{G}(\text{Sh})$ satisfies
\begin{align}
  \cot\left(\frac{\sqrt{\mathcal{G}(\text{Sh})}}{2}\right) = \frac{2\sqrt{\mathcal{G}(\text{Sh})}}{\text{Sh}}\,.\label{eq:gu}
\end{align}
The Sherwood number $\text{Sh}$ is a dimensionless number in fluid mechanics that represents the ratio of convective mass transfer over diffusive mass transport \cite{Sherwood}. It is the direct analog of the Nusselt number in heat transfer. Even if it seems difficult to find an analytical expression of the function $\mathcal{G}(\text{Sh})$ for arbitrary values of $\text{Sh}$, we present the asymptotic behavior of $\mathcal{G}(\text{Sh})$ for $\text{Sh}\to 0$ and $\text{Sh}\to\infty$ in the remaining of this section.

\begin{figure}[t]
\subfloat{%
 \includegraphics[width=0.5\textwidth]{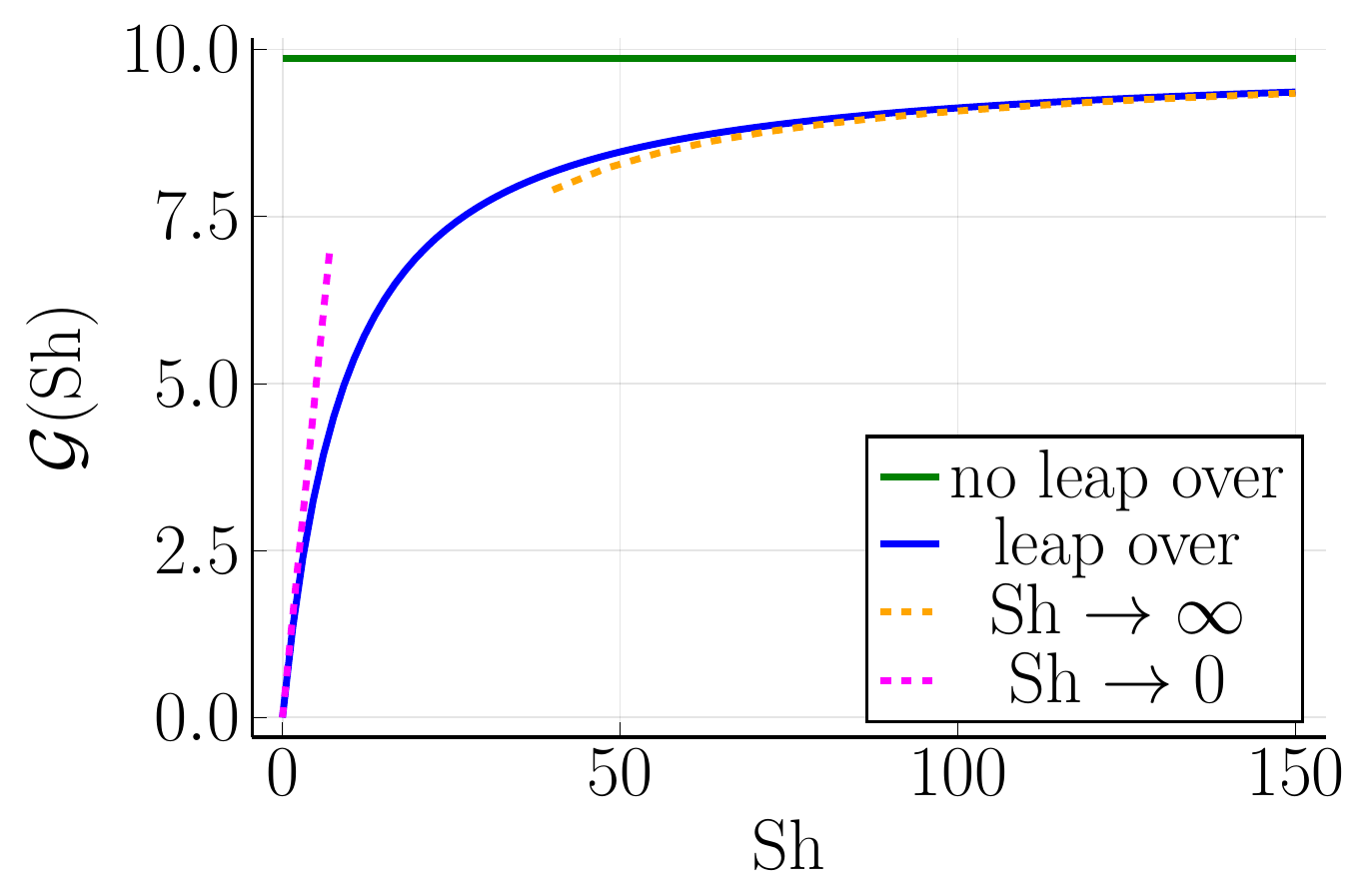}%
}\hfill
\subfloat{%
  \includegraphics[width=0.5\textwidth]{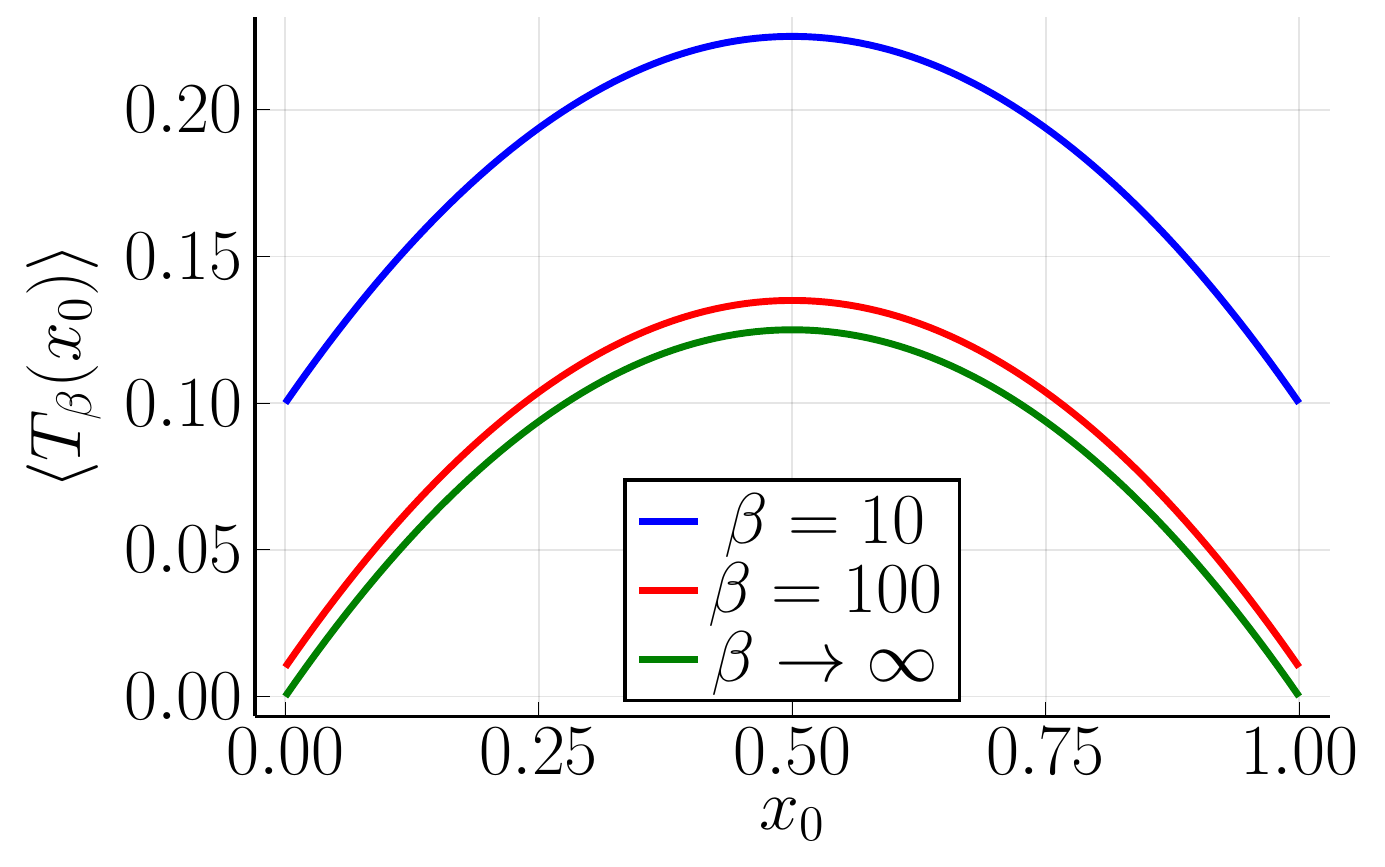}%
}\hfill
\caption{\textbf{Left panel:} Scaling function $\mathcal{G}(\text{Sh})$ of the decay rate $\alpha(\beta,L) =\lim_{t\rightarrow \infty} -\frac{\ln S_{\beta,L}(t\,|\,x_0) }{t}$  for a diffusive particle with periodically distributed point absorbers as a function of the Sherwood number $\text{Sh}=\frac{\beta L}{D}$. \textbf{Right panel:} Mean first-passage time $\langle T_{\beta,L}(x_0)\rangle $ of a Brownian motion with periodically distributed point absorbers as a function of the initial position $x_0$ with $L=1$ and $D=1$. Different curves are represented as a function of the absorption rate $\beta$. The case $\beta\rightarrow \infty$ corresponds to the case of a Brownian motion on an interval of length $L$ with absorbing boundaries.  }
\label{fig:bm}
\end{figure}

\paragraph{Limit of low Sherwood number} In the limit $\text{Sh}\to 0$, we find that the solution of the transcendental equation (\ref{eq:alphascal}) is
\begin{align}
  \mathcal{G}(\text{Sh}) \sim \text{Sh}-\frac{\text{Sh}^2}{12}\,,\quad \text{Sh}\rightarrow 0\,,\label{eq:beta0}
\end{align}
where we used that $\cot x=\frac 1 x -\frac x 3 +o(x)$. This asymptotic expansion exhibits a similar behavior as in the discrete result (\ref{eq:Salg}). Note that this result is also in agreement with equation (4.15) in \cite{Krapivsky2014} where the authors consider a particle on discrete ring with absorbing sites and nearest neighbour hopping dynamics. 

\paragraph{Limit of high Sherwood number} In the limit $\text{Sh}\to \infty$, we expect the decay rate to converge to a constant. By performing a $1/\text{Sh}$ expansion in the transcendental equation, we find
\begin{align}
 \mathcal{G}(\text{Sh}) \sim  \pi^2-\frac{8\pi^2}{\text{Sh} }\,,\quad \text{Sh}\rightarrow \infty\,,\label{eq:betainf}
\end{align}
where we recognise the leading order term as the decay rate of the survival probability of a diffusive particle in an interval of size $L$ with absorbing boundaries (see for instance equation 2.2.4 in \cite{rednerGuide}). The exact decay rate and the asymptotic results are shown in the left panel in figure \ref{fig:bm}.

\subsubsection{Mean first-passage time} The mean first-passage time can be directly extracted from the Laplace transform of the survival probability (\ref{eq:solS}) by setting $s=0$ which gives 
\begin{align}
 \langle T_\beta(x_0)\rangle = \frac{x_0\left(L-x_0\right)}{2D} + \frac{L}{\beta}\,.\label{eq:avgTc}
\end{align}
This result is consistent with the discrete-time result and can be directly obtained from it as we show in \ref{app:der diff lim}. The mean first-passage time is displayed as a function of the initial position in the right panel in figure \ref{fig:bm}. In the expression (\ref{eq:avgTc}), we see that in the limit $\beta\rightarrow 0$, the mean first-passage time diverges as the traps are no more absorbing. Alternatively, in the limit $\beta\rightarrow \infty$, we recover the mean first-passage time for a diffusive particle in a finite interval with absorbing boundaries (see for instance 2.2.17 in \cite{rednerGuide}).

\subsection{Diffusion with uniform absorption}\label{sec:diffUA}
In this section, we consider the diffusive limit 
\begin{align}
  \sigma\rightarrow 0\,,\quad \tau \rightarrow 0\,,\quad \ell\rightarrow 0\,,\quad L\rightarrow 0\,,\quad \text{with}\quad D\coloneqq \frac{\sigma^2}{2\tau}\,,\quad \alpha \coloneqq\frac{\ell}{\tau L}\quad \text{fixed}.\label{eq:diff2}
\end{align}
As in the previous section, we start from (\ref{eq:fordx}) and perform an expansion up to second order in $\eta$ to obtain (\ref{eq:fordd}). In the limit of small $\ell$ and $L$, the first term in brackets in (\ref{eq:fordd}) becomes
\begin{align}
 \sum_{m=-\infty}^{\infty} \int_{m\,(L+\ell)-x}^{m\,(L+\ell)+L-x} d\eta\, f(\eta) &\sim 1 -\sum_{m=-\infty}^{\infty}\int_{m(L+\ell)-x-\ell}^{m(L+\ell)-x} d\eta\, f(\eta) \,,\nonumber\\
  &\sim  1 -\ell\sum_{m=-\infty}^{\infty} f(m\,L-x)\,,\nonumber\\
  &\sim  1 -\ell\int_{-\infty}^{\infty}dm f(m\,L-x)\,,\nonumber\\
  &\sim  1 -\frac{\ell}{L}\,,\quad \ell\to 0\,,\label{eq:approxD3}
\end{align}
where we used the normalisation $\int_{-\infty}^{\infty} d\eta\, f(\eta)=1$ in the first line, the fact that $\ell$ is small in the second line, and that $L$ is small in the third line.  The term in the second bracket in (\ref{eq:fordd}) tends towards the variance of the jump distribution as in the previous section. Inserting the expansions of the two terms in brackets (\ref{eq:approxD2}) and (\ref{eq:approxD3}), and going from the number of steps $n$ to a continuous time $t$, we find the diffusion equation with uniform absorption on the real line:
\begin{align}
  \partial_t S_{\alpha}(t\,|\,\theta) = D \partial_{xx} S_{\alpha}(t\,|\,x)  - \alpha\, S_{\alpha}(t\,|\,x) \,,\label{eq:diffab}
\end{align}
 where $D$ and $\alpha$ are the diffusion coefficient and absorption coefficient respectively. The solution to the diffusion equation with uniform absorption on the real line with the initial condition $S_\alpha(t=0\,|\,x)=1$ is simply
\begin{align}
S(t\,|\,x) =e^{-\alpha t}\,,\qquad \forall x\in\mathbbm{R},\label{eq:solL02}
\end{align}
due to the fact that the absorption happens uniformly on the real line. 

\section{Summary and outlook}
In this work, we studied the survival probability of a random walk leaping over finite size and periodically distributed traps. We first studied a discrete-time random walk with a double-sided exponential jump distribution, for which we could compute the mean first-passage time and the survival probability explicitly. We found that the survival probability decays exponentially with a decay rate that depends in a non-trivial way on the length of the traps. In the limit of small traps, we found that the decay rate tends to the ratio of the length of the traps over the distance between them, which is interestingly the same result as in classical billiard problems. We then derived the mean first-passage time and the survival probability for continuous-time random walks with a power-law waiting time distribution. For such random walks, the survival probability decays algebraically with an exponent that is independent of the length of the traps. Finally, we derived the diffusive limit of our model and showed that, depending on the scaling, we obtain either diffusion with uniform absorption, or diffusion with periodically distributed point absorbers.

Going beyond the double-sided exponential jump distribution, it would be interesting to investigate the survival probability of random walks with arbitrary jump distributions leaping over traps. It was shown here that the decay rate of the survival probability, in the limit of small traps, is given by the ratio of the length of the traps over the distance between them. We expect this result to hold for more general jump distributions. Additionally, we would like to investigate to what extent is the additional term in the mean first-passage time, induced by the traps, independent of the initial position.

 Furthermore, it would be interesting to pursue further the connection with the classical billiard problems. In particular, it would be interesting to add a second trap, with a different length, and study the effect on the decay rate of the survival probability. The decay rate could possibly be given by the sum of two terms corresponding to each traps plus an ``interaction'' term between them, as it is the case for the classical billiard problem \cite{Bunimovich(2007)}. Another possible extension of this work would be to introduce disorder in location of the traps, for instance with L\'evy distributed distances between the traps \cite{Artuso18}. Finally, one could further develop the model by introducing moving traps that could provide a more accurate description of chemical phenomena where one has to take into account the motion of all the reactants \cite{Koza(1998), Sanchez(1998), Sanchez(1999)}, for instance when one species can annihilate upon collision or two chemical species combine together to form an inert product.

\section*{Acknowledgements}
BD warmly thanks Gr\'egory Schehr and Satya N. Majumdar for fruitful comments and feedback. BD gratefully acknowledges the financial support of the Luxembourg National
Research Fund (FNR) (App. ID 14548297). GP is deeply grateful to Roberto Artuso and acknowledges partial support from PRIN Research Project No. 2017S35EHN ``Regular and stochastic behavior in dynamical systems'' of the Italian Ministry of Education, University and Research (MIUR). Special thanks are also addressed to the summer school ``Fundamental Problems in Statistical Physics XV'' held in Bruneck (Italy), where this work was initiated.

\appendix
\section{Mean first-passage time for the double-sided exponential jump distribution} 
\label{app:mfpt}
In this appendix, we show how to obtain the mean first-passage time (\ref{eq:mpftsolx}) from the generating function of the survival probability (\ref{eq:dsesol}). To do so, we need to evaluate the generating function at $z=1$. We notice that the system of equations (\ref{eq:sysO}) becomes ill-defined exactly at $z=1$ as the determinant (\ref{eq:det}) vanishes. Given the right-hand side in (\ref{eq:sysO}) and the fact that the matrix $\mathbf{\Omega}(z,\varphi)$ in (\ref{eq:Omega}) behaves as
\begin{align} 
\left(
\begin{array}{cccc}
1 & 1 & -e^{\frac{2\sqrt{2}\pi}{\sigma_a}} & -e^{-\frac{2\sqrt{2}\pi}{\sigma_a}} \\
 \sqrt{1-z} & -\sqrt{1-z} &-\,e^{\frac{2\sqrt{2}\pi}{\sigma_a}} & e^{-\frac{2\sqrt{2}\pi}{\sigma_a}} \\
 1+ \frac{\sqrt{2}(2\pi-\varphi)\,\sqrt{1-z}}{\sigma_a} + \frac{2 (2\pi-\varphi)^2\,(1-z) }{\sigma_a^2}&1- \frac{\sqrt{2}(2\pi-\varphi)\,\sqrt{1-z}}{\sigma_a} + \frac{2(2\pi-\varphi)^2\,(1-z)}{\sigma_a^2} &  -e^{\frac{\sqrt{2}(2\pi-\varphi)}{\sigma_a}} &
- e^{-\frac{\sqrt{2}(2\pi-\varphi)}{\sigma_a}}   \\
 \sqrt{1-z}+\frac{\sqrt{2}(2\pi-\varphi)\,(1-z)}{\sigma_a} & -\sqrt{1-z}+\frac{\sqrt{2}(2\pi-\varphi)\,(1-z) }{\sigma_a}&
    -\,e^{\frac{\sqrt{2}(2\pi-\varphi)}{\sigma_a}}     & e^{-\frac{\sqrt{2}(2\pi-\varphi)}{\sigma_a}}
\end{array} 
\right) \label{eq:Omegae}
\end{align}
 when $z\to 1$, we seek for a solution of the form:
\begin{align}
   \left(
\begin{array}{c}
 A(z,\varphi)\\
 B(z,\varphi)\\
 C(z,\varphi)\\
 D(z,\varphi)
\end{array} 
\right) \sim \left(
\begin{array}{c}
 \frac{a_1}{1-z} + \frac{a_2}{\sqrt{1-z}} + a_3\\
  \frac{b_1}{1-z} + \frac{b_2}{\sqrt{1-z}} + b_3\\
 c_1\\
 d_1
\end{array} 
\right)\,, \quad z\rightarrow 1\,,
\label{eq:ABCDe}
\end{align}
where $a_1$, $a_2$, $a_3$, $b_1$, $b_2$, $b_3$, $c_1$ and $d_1$ are coefficient to be determined. We solve the system at orders $O[(1-z)^{-1}]$, $O[(1-z)^{-1/2}]$ and $O(1)$.
\paragraph{Order $O[(1-z)^{-1}]$} At this order, the only non-trivial equation to solve is:
\begin{align}
  a_1 +b_1=-1\,.\label{eq:order1}
\end{align}
\paragraph{Order $O[(1-z)^{-1/2}]$} At this order, the non-trivial equations to solve are:
\begin{subequations}
\begin{align}
  a_2 + b_2 &= 0\,,\\
  a_1 - b_1 &= 0\,.
\end{align}
\label{eq:order2}
\end{subequations}
\paragraph{Order $O(1)$} At this order, the non-trivial equations to solve are:
\begin{subequations}
\begin{align}
  a_3 + b_3  -e^{\frac{2\sqrt{2}\pi}{\sigma_a}} c_1 -e^{-\frac{2\sqrt{2}\pi}{\sigma_a}}d_1 &= 0\,,\\
  a_2 - b_2 -e^{\frac{2\sqrt{2}\pi}{\sigma_a}}c_1  + e^{-\frac{2\sqrt{2}\pi}{\sigma_a}}d_1 &= 0\,, \\
  a_3 + \frac{\sqrt{2}(2\pi-\varphi)}{\sigma_a} \, a_2 + \frac{2(2\pi-\varphi)^2}{\sigma_a^2} \, a_1 + b_3 - \frac{\sqrt{2}(2\pi-\varphi)}{\sigma_a}\, b_2 \qquad&\nonumber\\
+ \frac{2(2\pi-\varphi)^2}{\sigma_a^2} \, b_1 -e^{\frac{\sqrt{2}(2\pi-\varphi)}{\sigma_a}} c_1  - e^{-\frac{\sqrt{2}(2\pi-\varphi)}{\sigma_a}} d_1 &= 0\,,\\
 a_2-b_2 + \frac{\sqrt{2}(2\pi-\varphi)}{\sigma_a}\, a_1 + \frac{\sqrt{2}(2\pi-\varphi)}{\sigma_a}\, b_1 - e^{\frac{\sqrt{2}(2\pi-\varphi)}{\sigma_a}} c_1  + e^{-\frac{\sqrt{2}(2\pi-\varphi)}{\sigma_a}} d_1&= 0\,.
\end{align}
\label{eq:order3}
\end{subequations}
Combining the equations (\ref{eq:order1}), (\ref{eq:order2}) and (\ref{eq:order3}), we find
\begin{subequations}
\begin{align}
  a_1 &= b_1 = -\frac{1}{\sigma_a^2}\,,\\
   a_2 &= -b_2 = \frac{2\pi-\varphi}{2\sigma_a^2}\,,\\
   a_3 + b_3  &=\frac{(2\pi-\varphi)}{\sqrt{2}\sigma_a}\,\coth\left(\frac{\varphi}{\sqrt{2}\sigma_a}\right)\,,\\
   c_1 &= \frac{2\pi-\varphi}{2 \sigma_a\,e^{\frac{2\sqrt{2}\pi}{\sigma_a}}}\left[ \frac1{\sqrt{2}}\coth\left(\frac{\varphi}{\sqrt{2}\sigma_a}\right)+\frac 1{\sigma_a}\right] \,,\\
   d_1 &= \frac{(2\pi-\varphi)\,e^{\frac{2\sqrt{2}\pi}{\sigma_a}}}{2 \sigma_a}\left[ \frac1{\sqrt{2}}\coth\left(\frac{\varphi}{\sqrt{2}\sigma_a}\right)-\frac 1{\sigma_a}\right] \,.
\end{align}
\label{eq:solabcd}
\end{subequations}
Inserting the expansion of the integration coefficients (\ref{eq:ABCDe}) in the generating function (\ref{eq:dsesol}) and expanding to appropriate order, we get
\begin{align}
   q_{\varphi}(z,\theta) \sim \left\{\begin{array}{ll}
 \frac{1+a_1+b_1}{1-z}+\frac{a_2+b_2+\theta(a_1-b_1)}{\sqrt{1-z}}+\frac{a_1+b_1}{2}\,\theta^2 +(a_2-b_2)\,\theta+a_3+b_3\,,& 0\leq\theta\leq 2\pi-\varphi\,,\\[1em]
  c_1 \,e^{\theta}+ d_1\, e^{-\theta}\,,& 2\pi-\varphi<\theta<2\pi\,,
  \end{array}\right. \label{eq:dsesolea}
\end{align}
for $z\to 1\,.$ Inserting the solution for the integration constants (\ref{eq:solabcd}), we find:
\begin{align}
   q_{\varphi}(z,\theta) \sim \left\{\begin{array}{ll}
 -\frac{\theta^2}{\sigma_a^2} + \frac{\theta(2\pi-\varphi)}{\sigma_a^2}+\frac{(2\pi-\varphi)}{\sqrt{2}\sigma_a}\,\coth\left(\frac{\varphi}{\sqrt{2}\sigma_a}\right)\,,& 0\leq\theta\leq 2\pi-\varphi\,,\\[1em]
   c_1 \,e^{\theta}+  d_1\, e^{-\theta}\,,& 2\pi-\varphi<\theta<2\pi\,,
  \end{array}\right. \qquad z\rightarrow 1\,.\label{eq:dsesole}
\end{align}
Hence, we notice that the divergent terms cancel and we recover the expression of the mean first-passage time \eqref{eq:mpftsol} displayed in the main text.

\section{Mean first-passage time for a double-sided exponential jump distribution with no leap over}\label{app:nojump}
In this appendix, we compute the mean first-passage time for a double-sided exponential jump distribution on an interval $[0,L]$ with absorbing boundaries. The mean first-passage time $\langle T_{L}(x) \rangle_{\text{no leap over}}$, given the initial position of the random walk $x$, satisfies the following recursive relation
\begin{align}
  \langle T_{L}(x) \rangle_{\text{no leap over}} = 1 + \int_0^L dy f(y-x)   \langle T_{L}(y) \rangle_{\text{no leap over}}\,,\label{eq:Tnoje}
\end{align}
where $f(\eta)$ is the  double-sided exponential jump distribution (\ref{eq:dse}). Taking twice a position derivative of (\ref{eq:Tnoje}) and using that $\sigma^2/2 f''(\eta)=f(\eta)-\delta(\eta)$, we get
\begin{align}
  \frac{\sigma^2}{2}\,\partial_{xx} \langle T_{L}(x) \rangle_{\text{no leap over}} =  \langle T_{L}(x) \rangle_{\text{no leap over}}-1 - \Theta(x)\Theta(L-x) \langle T_{L}(x) \rangle_{\text{no leap over}}\,.\label{eq:diffTnjp}
\end{align}
This equation can be solved on the three intervals $]-\infty,0[$, $[0,L]$ and $]L,\infty[$. The general solution is
\begin{align}
   \langle T_{L}(x) \rangle_{\text{no leap over}} = \left\{\begin{array}{ll}
     A(L)\,e^{\frac{\sqrt{2}x}{\sigma}} + B(L)\,e^{-\frac{\sqrt{2}x}{\sigma}}\,, \qquad& x < 0\,,\\
     -\frac{x^2}{\sigma^2} + \frac{C(L) x}{\sqrt{2}\sigma} + D(L)\,,\qquad& 0\leq x\leq L\,,\\
      E(L)\,e^{\frac{\sqrt{2}x}{\sigma}} + F(L)\,e^{-\frac{\sqrt{2}x}{\sigma}}\,, \qquad& x > L\,.
   \end{array}\right.\label{eq:tnpgsol}
\end{align}
By requiring that the solution vanishes at $x\rightarrow \pm \infty$, we find $B(L)=0$ and $E(L)=0$. By further requiring the continuity of the solution and its derivative at $x=0$ and $x=L$, we find
\begin{align}
A(L)= \frac{L}{\sqrt{2} \sigma }\,,\quad  C(L)= \frac{\sqrt{2}L}{\sigma }\,,\quad D(L)= \frac{L}{\sqrt{2} \sigma }\,,\quad F(L)= \frac{L e^{\frac{\sqrt{2} L}{\sigma  }}}{\sqrt{2} \sigma }\,,\label{eq:constintnj}
\end{align}
which recovers the expression (\ref{eq:mpftsolx_nojumps}) in the main text.

\section{Asymptotic behavior of the decay rate $\alpha(\ell,L)$ of the survival probability for $L\to 0$ and $L\to \infty$}
\label{app:L0I}

In this appendix, we derive the asymptotic behavior of the decay rate $\alpha(\ell,L)$, satisfying the transcendental equation (\ref{eq:detzero}), in the limit of $L\to 0$ and $L\to \infty$. 
\subsection{Limit $L\to 0$} In this limit, we expect the decay rate to diverge as the traps become infinitely close to each others. The transcendental equation (\ref{eq:detzero}) can be expanded to first order in $L$, which gives the equation
\begin{align}
   (e^{\alpha(\ell,L)}-2)\sinh\left(\frac{\sqrt{2}\ell}{\sigma}\right) \frac{\sqrt{2}\,L}{\sigma}+2 \left[1-\cosh\left(\frac{\sqrt{2}\ell}{\sigma}\right) \right] = 0\,,\qquad L\to 0\,.\label{eq:detzero01}
   \end{align}
Solving it for $\alpha(\ell,L)$ and taking the limit $L\rightarrow 0$ gives
\begin{align}
  \alpha(\ell,L) \sim \ln\left(\frac{1}{L}\right)\,,\qquad L\to 0\,,\label{eq:alpha0}
\end{align}
which recovers expression (\ref{eq:alpha0m}) in the main text.

\subsection{Limit $L\to \infty$} In the limit $L\to \infty$, we expect the decay rate to vanish as the distance between the traps tends to infinity. If we look for an expansion $ \alpha(\ell,L) \sim \frac{b_1(\ell)}{L^2}$ for $L\rightarrow\infty$, we obtain from the transcendental equation that $b_1(\ell)=\pi^2\sigma^2/2$, which gives
\begin{align}
  \alpha(\ell,L) \sim \frac{\pi^2\sigma^2}{2L^2}\,,\qquad L\to \infty\,,\label{eq:alphaLi}
\end{align}
which recovers expression (\ref{eq:alphaLim}) in the main text.

\section{Laplace transform of the survival probability and the mean first-passage time for CTRWs} 
\label{app:LapCTRW}
In this appendix, we derive the exact Laplace transform of the survival probability and the mean first-passage time in the CTRW framework.
\subsection{Survival probability}
The survival probability $S_{\varphi}(t\,|\, \theta)$ satisfies the following recurrence relation
\begin{align}
S_{\varphi}(t\,|\, \theta)&=\int_0^t dt'\,\psi(t')\, \int_0^{2\pi-\varphi}d\theta ' F(\theta '-\theta) S_{\varphi}(t-t'\,|\,\theta')\,, \label{eq:SurRec}
\end{align}
where $\psi(t)$ is the waiting time distribution (\ref{eq:psit}) and $F(\theta)$ is the periodised jump distribution (\ref{eq:Fperdes}). Note that (\ref{eq:SurRec}) boils down to \eqref{eq:Srec} for $\psi(t)=\delta(t-1)$. In the Laplace domain, it reads
\begin{align}
\tilde{S}_{\varphi}(s\,|\,\theta)=\frac{1-\tilde{\psi}(s)}s +\tilde{\psi}(s)\int_0^{2\pi-\varphi}d\theta ' \, F(\theta'-\theta)\, \tilde{S}_{\varphi}(s\,|\,\theta')\,. \label{eq:SurRecLap}
\end{align}
By applying the same method as in the discrete-time random walk, we find that it follows the differential equation
\begin{align}
\frac{\sigma_a^2}2\partial_{\theta\theta}\tilde{S}_{\varphi}(s\,|\,\theta)&=\tilde{S}_{\varphi}(s\,|\,\theta)-\frac{1-\tilde{\psi}(s)}s-\tilde{\psi}(s)\tilde{S}_{\varphi}(s\,|\,\theta)\Theta(2\pi-\varphi-\theta)\,.
\label{eq:SurDiffLap}
\end{align}
We solve the differential equation (\ref{eq:SurDiffLap}) on the two intervals $[0,2\pi-\varphi]$ and $]2\pi-\varphi,2\pi[$. The general solution is
\begin{align}
\tilde{S}_{\varphi}(s\,|\,\theta)-\frac 1 s = \begin{cases}
A(s,\,\varphi)\, e^{\frac{\sqrt{2}\sqrt{1-\tilde{\psi}(s)}\theta}{\sigma_a}}+B(s,\,\varphi)\, e^{-\frac{\sqrt{2}\sqrt{1-\tilde{\psi}(s)}\theta}{\sigma_a}}\,, \qquad & 0\leq \theta\leq 2\pi-\varphi\,,\\
C(s,\,\varphi)\, e^{\frac{\sqrt{2}\theta}{\sigma_a}}+D(s,\,\varphi)\, e^{-\frac{\sqrt{2}\theta}{\sigma_a}}-\frac{\tilde{\psi}(s)} s\,,\qquad & 2\pi-\varphi<\theta<2\pi \,.
\end{cases}
\label{eq:SurSolLapShift}
\end{align}
To fix the integration constants $A(s,\varphi)$, $B(s,\varphi)$, $C(s,\varphi)$, $D(s,\varphi)$, we impose the periodic boundary conditions of the solution and its derivative and we match the solution and its derivative at $\theta=2\pi-\varphi$. This gives the following linear system
\begin{align}
\mathbf{\Omega}(s,\varphi)
\left(
\begin{array}{c}
 A(s,\varphi)\\
 B(s,\varphi)\\
 C(s,\varphi)\\
 D(s,\varphi)
\end{array} 
\right) =   \left(
\begin{array}{c}
 -\frac{\tilde{\psi}(s)} s\\
0\\
-\frac{\tilde{\psi}(s)} s\\
0
\end{array} 
\right)\,,\label{eq:ABCDsolSurv}
\end{align}
where the matrix $\mathbf{\Omega}(s,\varphi)$ is given by
\begin{align}
\mathbf{\Omega}(s,\varphi)=
 \left(
\begin{array}{cccc}
 1 & 1 & -e^{\frac{2\sqrt{2}\pi}{\sigma_a}} & -e^{-\frac{2\sqrt{2}\pi}{\sigma_a}} \\
 \sqrt{1-\tilde{\psi}(s)} & -\sqrt{1-\tilde{\psi}(s)} &-\,e^{\frac{2\sqrt{2}\pi}{\sigma_a}} & e^{-\frac{2\sqrt{2}\pi}{\sigma_a}} \\
 e^{\frac{\sqrt{2}\sqrt{1-z}(2\pi-\varphi)}{\sigma_a}} & e^{-\frac{\sqrt{2}\sqrt{1-z}(2\pi-\varphi)}{\sigma_a}} & -e^{\frac{\sqrt{2}(2\pi-\varphi)}{\sigma_a}} &
- e^{-\frac{\sqrt{2}(2\pi-\varphi)}{\sigma_a}}   \\
\sqrt{1-\tilde{\psi}(s)}\,e^{\frac{\sqrt{2}\sqrt{1-\tilde{\psi}(s)}(2\pi-\varphi)}{\sigma_a}}& -\sqrt{1-z}\,e^{\frac{-\sqrt{2}(2\pi-\varphi)}{\sigma_a}}  & -\,e^{\frac{\sqrt{2}(2\pi-\varphi)}{\sigma_a}}
    & e^{-\frac{\sqrt{2}(2\pi-\varphi)}{\sigma_a}}\\
\end{array} 
\right)
\,.\label{eq:ABCDmatSurv}
\end{align}
The exact solution is too long to be displayed and the relevant information is given in the main text by means of asymptotic estimates.

\subsection{Mean first-passage time}
The mean first-passage time follows the recursive relation
\begin{align}
\langle T_{\varphi}(\theta)\rangle_{\text{CTRW}} = \langle \tau \rangle +\int_0^{2\pi-\varphi}d\theta'\,F(\theta'-\theta) \,\langle T_{\varphi}(\theta')\rangle_{\text{CTRW}}\,,\label{eq:mfpt_rec}
\end{align}
where $\langle \tau \rangle \coloneqq\int_0^{\infty} dt \, t\, \psi(t)$ is the mean waiting time. By applying the same method as in the discrete-time random walk, we find that it follows the differential equation 
\begin{align}
 \frac{\sigma_a^2}2 \partial_{\theta\theta}\,\langle T_{\varphi}(\theta)\rangle_{\text{CTRW}} =   \langle T_{\varphi}(\theta)\rangle_{\text{CTRW}}-\langle \tau \rangle - \,\Theta(2\pi-\varphi-\theta) \langle T_{\varphi}(\theta)\rangle_{\text{CTRW}}\,,\quad 0\leq \theta <2\pi\,,\label{eq:diffT}
\end{align}
whose solution is given in (\ref{eq:mpftsolgen}).

\section{Direct derivation of diffusive results from the discrete results} 
\label{app:der diff lim}
In this appendix, we derive the diffusive results on the survival probability and the mean first-passage time from the discrete results.
\subsection{Mean first-passage time}
We start from \eqref{eq:mpftsol} and multiply by the time step $\tau$ of the diffusive limit:
\begin{align}
\langle T_{\beta}(\theta) \rangle &= \tau  + \frac{\tau x_0(L-x_0)}{\sigma^2}+\frac{\tau L}{\sqrt{2}\sigma}\,\coth\left(\frac{\ell}{\sqrt{2}\sigma}\right)\,.\label{eq:tappba}
\end{align}
In the diffusive limit $\ell\to 0$, using the asymptotic approximation $\coth(x)\sim \frac 1 x $ when $x\to 0$, we directly obtain the diffusive result (\ref{eq:avgTc}).

\subsection{Decay rate}
We start from (\ref{eq:detzero}), set $\alpha(\ell,L)=\alpha(\beta,L)\,\tau$ and take the diffusive limit $\ell \rightarrow 0$, $\tau\rightarrow 0$, $\sigma\rightarrow 0$, while fixing $\ell/\tau$ and $\sigma^2/\tau$, which gives:
\begin{align} &- \frac{\sqrt{2}\ell}{\sigma}\sin \left(\frac{\sqrt{2}\,L\sqrt{\alpha(\beta,L)\tau}}{\sigma}\right) +2 \sqrt{\alpha(\beta,L)\tau}\left[1- \cos \left(\frac{\sqrt{2}\,L \sqrt{\alpha(\beta,L)\tau}}{\sigma}\right)\right] = 0\,.\label{eq:detzeroA}
  \end{align}
Identifying $\beta=\ell/\tau$ and $D=\sigma^2/2\tau$ gives
\begin{align}
  -\frac{\beta}{2\sqrt{\alpha(\beta,L)D}}\sin\left(L\sqrt{\frac{\alpha(\beta,L)}{D}}\right) +1 - \cos\left(L\sqrt{\frac{\alpha(\beta,L)}{D}}\right) = 0\,.\label{eq:detzeroA2}
\end{align}
Finally, using the trigonometric identities $\sin(2a)=2\sin(a)\cos(a)$ and $\cos(2a)=1-2\sin^2(a)$, we recover (\ref{eq:transc}).

\bibliographystyle{iopart-num}

\providecommand{\newblock}{}

\end{document}